\documentclass[aps,prl,reprint,superscriptaddress,secnumarabic]{revtex4-2}

\usepackage{graphicx}
\usepackage[version=3]{mhchem} 
\usepackage{verbatim}
\usepackage[utf8]{inputenc}
\usepackage{siunitx}
\usepackage{float}
\usepackage{comment}
\usepackage{textgreek}
\usepackage{sidecap} 

\usepackage{xcolor}

\begin{document}



\title{Geometry-induced spin-filtering in photoemission maps from WTe$_2$ surface states}

\author{Tristan Heider}
\affiliation{Peter Gr{\"u}nberg Institut (PGI-6), Forschungszentrum J{\"u}lich GmbH,
52428 J{\"u}lich, Germany}

\author{Gustav Bihlmayer}
\affiliation{Peter Gr{\"u}nberg Institut (PGI-1) and Institute for Advanced Simulation (IAS-1), Forschungszentrum J{\"u}lich and JARA, 52428 J{\"u}lich, Germany}

\author{Jakub Schusser}
\affiliation{New Technologies-Research Center, University of West Bohemia, 30614 Pilsen, Czech Republic}
\affiliation{Experimentelle Physik VII and Würzburg-Dresden Cluster of Excellence ct.qmat, Universität Würzburg, Würzburg, Germany}

\author{Friedrich Reinert}
\affiliation{Experimentelle Physik VII and Würzburg-Dresden Cluster of Excellence ct.qmat, Universität Würzburg, Würzburg, Germany}

\author{Jan Min\'ar}
\affiliation{New Technologies-Research Center, University of West Bohemia, 30614 Pilsen, Czech Republic}

\author{Stefan Bl\"ugel}
\affiliation{Peter Gr{\"u}nberg Institut (PGI-1) and Institute for Advanced Simulation (IAS-1), Forschungszentrum J{\"u}lich and JARA, 52428 J{\"u}lich, Germany}

\author{Claus M. Schneider}
\affiliation{Peter Gr{\"u}nberg Institut (PGI-6), Forschungszentrum J{\"u}lich GmbH,
52428 J{\"u}lich, Germany}
\affiliation{Fakult{\"a}t f{\"u}r Physik, Universit{\"a}t Duisburg-Essen, 47048 Duisburg, Germany}
\affiliation{Physics Department, University of California, Davis, CA 95616, USA}

\author{Lukasz Plucinski}
\email{l.plucinski@fz-juelich.de}
\affiliation{Peter Gr{\"u}nberg Institut (PGI-6), Forschungszentrum J{\"u}lich GmbH,
52428 J{\"u}lich, Germany}

\date{\today}

\begin{abstract}
We demonstrate that an important quantum material WTe$_2$ exhibits a new type of geometry-induced spin-filtering effect in photoemission, stemming from low symmetry that is responsible for its exotic transport properties. Through the laser-driven spin-polarized angle-resolved photoemission Fermi surface mapping, we showcase highly asymmetric spin textures of electrons photoemitted from the surface states of WTe$_2$. Such asymmetries are not present in the initial state spin textures, which are bound by the time-reversal and crystal lattice mirror plane symmetries. The findings are reproduced qualitatively by theoretical modeling within the one-step model photoemission formalism. The effect could be understood within the free-electron final state model as an interference due to emission from different atomic sites. The observed effect is a manifestation of time-reversal symmetry breaking of the initial state in the photoemission process, and as such it cannot be eliminated, but only its magnitude influenced, by special experimental geometries.
\end{abstract}

\maketitle



\begin{figure*}	
\centering
		\includegraphics[width=1\textwidth]{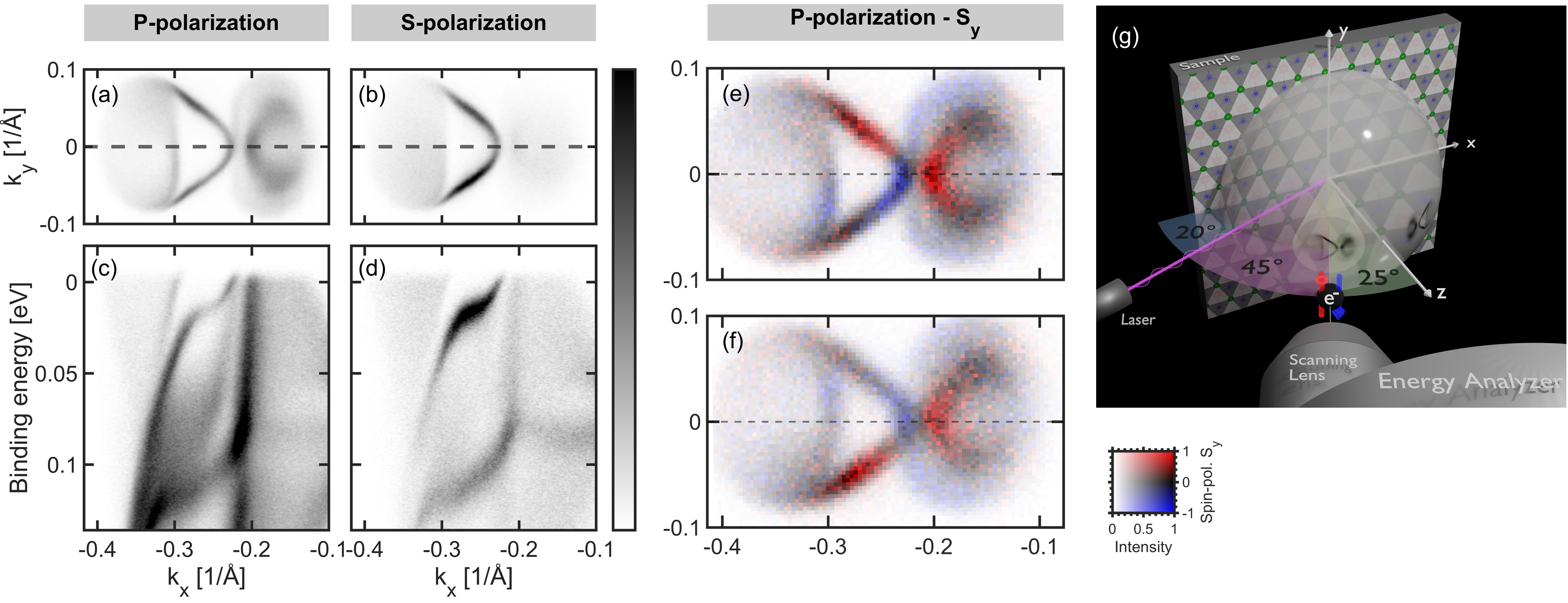}
	\caption{(a-b) Spin-integrated laser ARPES Fermi surface maps measured with $p$- and $s$-polarized light. (c-d) Corresponding energy dispersions $E(k_x)$ for $k_y=0$ as indicated by the dashed lines in (a) and (b). (e-f) Experimental laser-SARPES $57 \times 89$ pixel Fermi surface maps taken at two nearby spots on the same cleave. The false color scale refers to the spin polarization $S_{fy}$ in the ensemble of the photoemitted electrons. (g) Schematic geometry of the SARPES experiment. The maps were measured using the lens deflector system collecting the emission angles indicated by the yellow cone with the sample rotated by $\theta = 25^\circ$ with respect to the lens axis using $p$-polarized light and probing the spin along the $y$ axis, as defined in (g).}
	\label{fig:spinARPES}
\end{figure*}


{\it Introduction.---}WTe$_2$ is a semi-metallic two-dimensional (2D) quantum material that exhibits a non-saturating magnetoresistance up to 60 T \cite{Ali2014}. It has been debated, whether the bulk electron and hole pockets in WTe$_2$ slightly overlap leading to the Weyl type-II topological phase \cite{Soluyanov2015,Ruessmann2018}, with a conjecture that the surface electronic structure would be virtually indistinguishable for a topological and trivial phases \cite{Bruno2016}. When thinned down to a monolayer, WTe$_2$ enables the realization of high-temperature quantum Hall phases \cite{Tang2017} and gate-controlled superconductivity \cite{Fatemi2018,Sajadi2018}. In non-magnetic systems the first-order Hall response vanishes at zero magnetic field due to symmetry arguments. However, the second-order correction leads to the recently discovered \cite{Sodemann2015,Ma2019,Kang2019,Du2021} non-linear Hall effect (NLHE) in systems of reduced symmetry. Few-layer WTe$_2$ has been the first system in which the NLHE has been demonstrated \cite{Kang2019,Ma2019} due to the presence of a single mirror plane, and a related existence of polar axes both in-plane and out-of-plane of the layers.

Early high resolution angle-resolved photoemission (ARPES) on WTe$_2$ \cite{Pletikosic2014,Wu2015} have focused on imaging bulk electron and hole pockets with the motivation to explain the non-saturating magnetoresistance, pointing to a possible full charge compensation at lower temperatures. The prediction of possible type-II Weyl states in WTe$_2$ \cite{Soluyanov2015} have further motivated the research on its electronic structure. The reduced symmetry of WTe$_2$ results in inequivalent top and bottom surfaces along the $c$-axis. The general shape of the experimental WTe$_2$ band dispersions over the scale of 2 eV below the Fermi level is well reproduced by {\it ab initio} calculations, however, as pointed out in an early study \cite{Pletikosic2014}, density functional theory (DFT) calculations are not able to quantitatively reproduce critical features of the WTe$_2$ electronic structure near the Fermi level. In particular, this concerns the positions of the electron and hole pockets and related sizes of their Fermi surfaces \cite{DiSante2017} which play a critical role in understanding the magnetoresistance properties \cite{Pletikosic2014,Pippard1989}. This lack of agreement, despite occurring on a small energy scale of only 50 meV, is surprising because WTe$_2$(001) surfaces are clearly very stable under UHV yielding quantitatively consistent results in high-resolution ARPES for the set of dispersive occupied features located near the Fermi level \cite{Pletikosic2014,Wu2015,Bruno2016,Wang2017,DiSante2017,Wu2017,Rossi2020}.

Using the newly-developed high resolution laser-driven spin-polarized ARPES (SARPES) spectrometer we demonstrate for the first time the spin texture of the Fermi level photoemission map measured with 6 eV photons. Previous SARPES studies only probed selected regions in the Brillouin zone (BZ) \cite{Feng2016,Fanciulli2020,Wan2022}. We further demonstrate that the symmetry of the ARPES spin texture reflects the WTe$_2$ surface symmetry with a single mirror plane present, and not the initial state spin symmetry. Therefore, we directly demonstrate that the ARPES photocurrent carries additional information beyond the initial band structure spin texture, due to what we call {\it the geometry-induced spin filtering effect}.


The results are analyzed by comparison to the dedicated theoretical DFT calculations. The results obtained within the linearized-augmented plane-wave (LAPW) scheme show that the initial state spin texture follows the expected axial vector transformation rules of a single mirror plane and time-reversal. Further one-step model photoemission calculations within the Korringa-Kohn-Rostoker (KKR) scheme reproduce the (broken) symmetry properties of the experimental data. Finally, within the free-electron final state ARPES model we identify the origin of the observed asymmetries as an interference of the emission from different atomic sites.



{\it Methods.---} The sample (Td-WTe$_2$ single crystal, space group $Pmn2_1$, purchased from {\it HQ graphene}) was glued to the molybdenum sample plate by a silver-epoxy. We used $p$- or $s$-polarized light from the \textit{LEOS Solutions} continuous wave laser with $h\nu=6.02$~eV focused down to $\sim 50$~$\mu$m, the \textit{MB Scientific} A1 hemispherical electron analyzer and the exchange-scattering \textit{Focus GmbH} FERRUM spin detector \cite{Escher2011}. Spectrometers based on a similar design are in operation at synchrotron light sources \cite{Okuda2008,Bigi2017}. The mirror plane of WTe$_2$ was aligned parallel to the entrance slit of the
A1 spectrometer. The mechano-electrostatic lens deflector system of A1 allows mapping the emission angle over approx. $\pm 15^\circ$ in both $k_x$ and $k_y$ directions, therefore allowing for point-by-point $k_x$ vs. $k_y$ mapping in the spin-polarized mode without rotating the sample. All measurements were carried out at $\sim 15$~K at the pressure in the analyzer chamber $<5\times 10^{-11}$~mbar.

The initial state band structure was calculated using DFT in the generalized gradient approximation~\cite{Perdew1996}. We use the full-potential LAPW method in film geometry as implemented in the FLEUR code~\cite{Fleur}. Photoemission calculations were performed within the one-step model formalism as implemented in the fully relativistic KKR method \cite{Braun2018,Fanciulli2020}.

Further details on methods are provided in the Section SI of the Supplemental Material.


\begin{figure}
    \centering
    \includegraphics[width=\columnwidth]{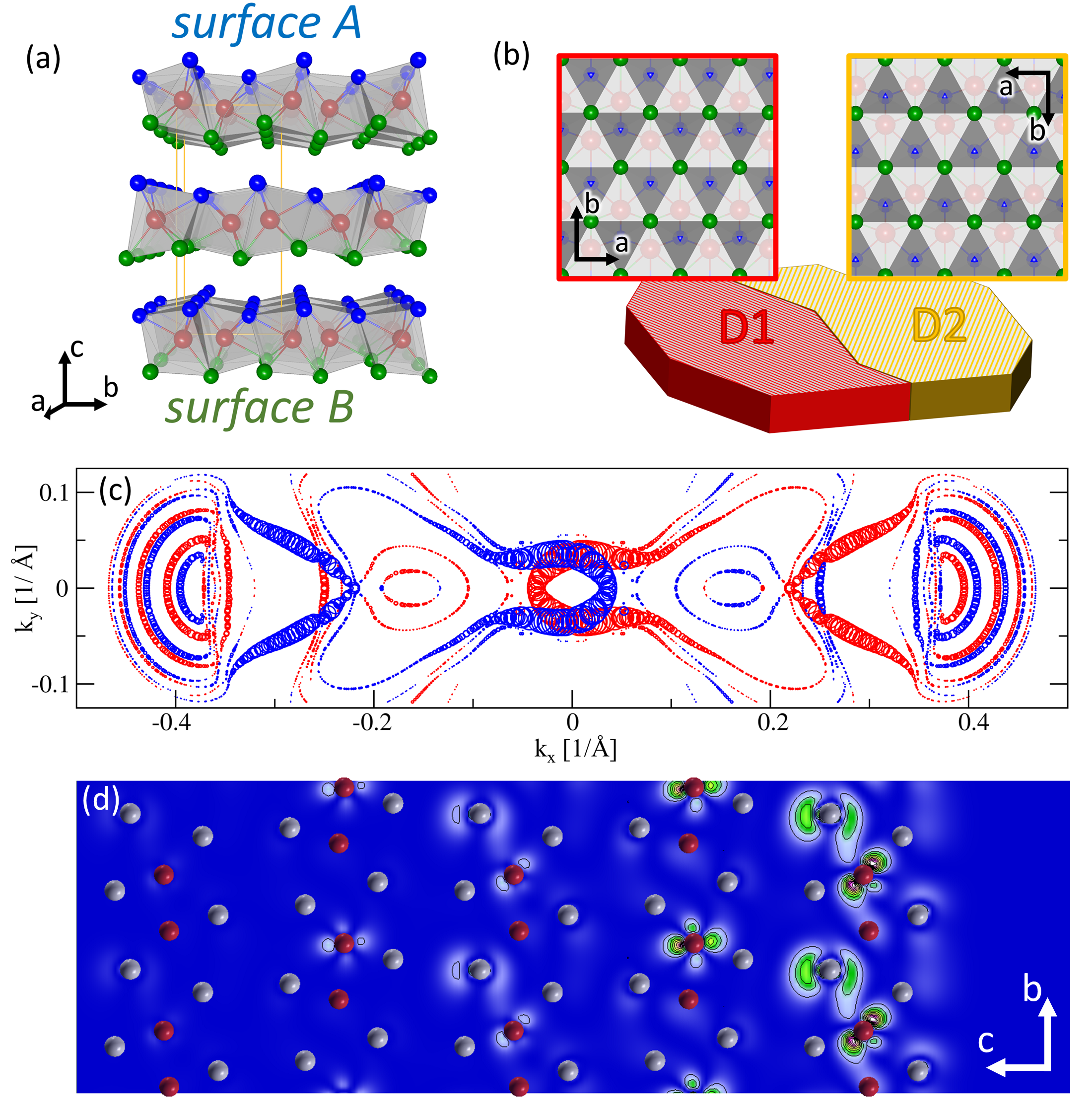}
    \caption{Crystal geometry and initial state spin textures. (a) The 3D impression of the WTe$_2$ crystal structure. (b) The probed surface with two possible terminations, which we label $D1$ and $D2$. (c) The $S_y$ component of the theoretical Fermi level spin texture, which is the same for both domains. The size of the symbols corresponds to the spin-polarization in the first layer (containing two formula units). (d) The surface state charge density in real space at $k_{x}=-0.3$\AA$^{-1}$ and initial energy $-25$ meV (see Supplemental Material Fig. S6).}
    \label{fig:domains}
\end{figure}


{\it Results.---}Figure \ref{fig:spinARPES}(a-d) shows the high-resolution ARPES maps measured with the 6 eV laser which are in quantitative agreement with previously published results \cite{Bruno2016,Wang2017}. Figure \ref{fig:spinARPES}(e-f) shows the spin-polarized Fermi surface maps with $p$-polarized light from two different spots on the sample in the off-normal geometry described in Fig. \ref{fig:spinARPES}(g), using the A1 lens deflector system to map the important portion of the BZ without rotating the sample. The use of the deflector system is critical to assure that the entire map is taken with the laser beam focused on precisely the same spot on the sample. Both maps show the familiar shape previously reported in Refs. \cite{Bruno2016,Wang2017}, however, their spin textures are different and highly asymmetric. Supplemental Material Fig. S4 shows SARPES spin texture taken with $s$-polarized light, where the spin polarization asymmetry, albeit smaller, is also observed.

The crystal lattice of WTe$_2$ is shown schematically in Fig. \ref{fig:domains}(a-b). The surface of WTe$_2$ exhibits a single $\mathcal M_x$ mirror plane, while the $y$ and $z$ directions are polar (see Section SII of the Supplemental Material and Ref. \cite{Weber2018}). By convention, the polarity along the $z$ axis has been defined as surface $A$ (or $top$) and $B$ (or $bottom$), as depicted in Fig. \ref{fig:domains}(a), with radically different Fermi contours measured on these surfaces \cite{Bruno2016,Rossi2020}. The shapes of Fermi surfaces from Fig. \ref{fig:spinARPES}(a-b,e-f) have been associated with the surface $B$.

The glide-reflection operation $(1/2+x,-y,1/2+z)$ of the $Pmn2_1$ space group \cite{Weber2018} shows that adjacent terraces, which we call $D1$ and $D2$, have reversed polarity along $y$, as shown explicitly in Fig. \ref{fig:domains}(b). Cleaved TMDCs typically exhibit macroscopically large terraces of the sizes of 100 $\mu$m or larger, and with our laser beamspot of 50 $\mu$m we can routinely address a single terrace, therefore we assume that the spectra in Fig. \ref{fig:spinARPES}(e-f) are taken on single terraces of reversed polarity.

Since WTe$_2$ is non-magnetic, the spin expectation value $\mathbf{S}_i$ of the initial state has to follow $\mathbf{S}_i(\mathbf{k}_i) = -\mathbf{S}_i(\mathbf{-k}_i)$. For the surface states, by combining this with the axial vector rules for the $\mathcal M_x$ mirror plane, one gets $S_{iy}(k_{ix},k_{iy}) = S_{iy}(k_{ix},-k_{iy})$, and our theoretical initial state spin texture for surface $B$ shown in Fig. \ref{fig:domains}(c) obeys this symmetry. However, this symmetry is broken in Fig. \ref{fig:spinARPES}(e) and in \ref{fig:spinARPES}(f). Already from the visual inspection in some portions of these SARPES maps $S_{fy}$ changes sign between $(k_{fx},k_{fy})$ and $(k_{fx},-k_{fy})$, while in other portions it does not. Here, ${\mathbf k}_f$ refers to the momentum and $S_{fy}$ to the $y$ component of the spin expectation value in the ensemble of photoemitted electrons measured by SARPES. We note that in ARPES, the component of ${\mathbf k}_f$ parallel to the surface is related to the off-normal emission angle $\theta$  and kinetic energy $E_{kin}$ by $k_{f_{||}} = (1/\hbar)\sqrt{2m E_{kin}} \sin{\theta}$ and the parallel momentum component is conserved in the photoemission process, i.e. ${\mathbf k}_{f||}  = {\mathbf k}_{i||}$.

A visual inspection suggests that Fig. \ref{fig:spinARPES}(e) can be transformed into Fig. \ref{fig:spinARPES}(f) by $S_{fy}(k_{fx},k_{fy}) \rightarrow S_{fy}(k_{fx},-k_{fy})$, which is indeed confirmed by a quantitative standard deviation analysis, see Section SI of the Supplemental Material. This can be explained by the symmetry operation $(1/2+x,-y,1/2+z)$ of the $Pmn2_1$ space group, assuming (e) and (f) are measured on adjacent terraces. Here the $1/2+z$ operation indicates switching into the adjacent layer/terrace. Reciprocal methods such as ARPES are insensitive to lateral shifts of the entire lattice and therefore the $1/2$ component in the $1/2+x$ operation can be ignored when considering the symmetries of ARPES. This means that SARPES maps from adjacent terraces are connected by the $\mathcal M_y$ mirror operation. Since in our geometry neither the $s$-polarized nor the $p$-polarized light is breaking the $\mathcal M_y$ mirror plane, this leads to $\mathcal M_y \{ S_{fy}^{D1}(k_{fx},k_{fy}) \} = S_{fy}^{D2}(k_{fx},-k_{fy})$ for our ARPES maps from adjacent terraces $D1$ and $D2$ (Fig. \ref{fig:domains}(b)).

Figure \ref{fig:domains}(d) shows that surface state on surface $B$ is primarily localized within the first two WTe$_2$ monolayers. One can also recognize a complex orbital structure of the surface state, with different orbital orientations contributing within the first and the second WTe$_2$ layers indicating radical breaking of the $\mathcal M_y$ mirror symmetry. Approximate shapes of $p_z$ orbitals on Te atoms and $d_{z^2}$ on W atoms can be recognized, however, they are not perfectly aligned along the $z$ axis.

\begin{figure}
    \centering
    \includegraphics[width=1\columnwidth]{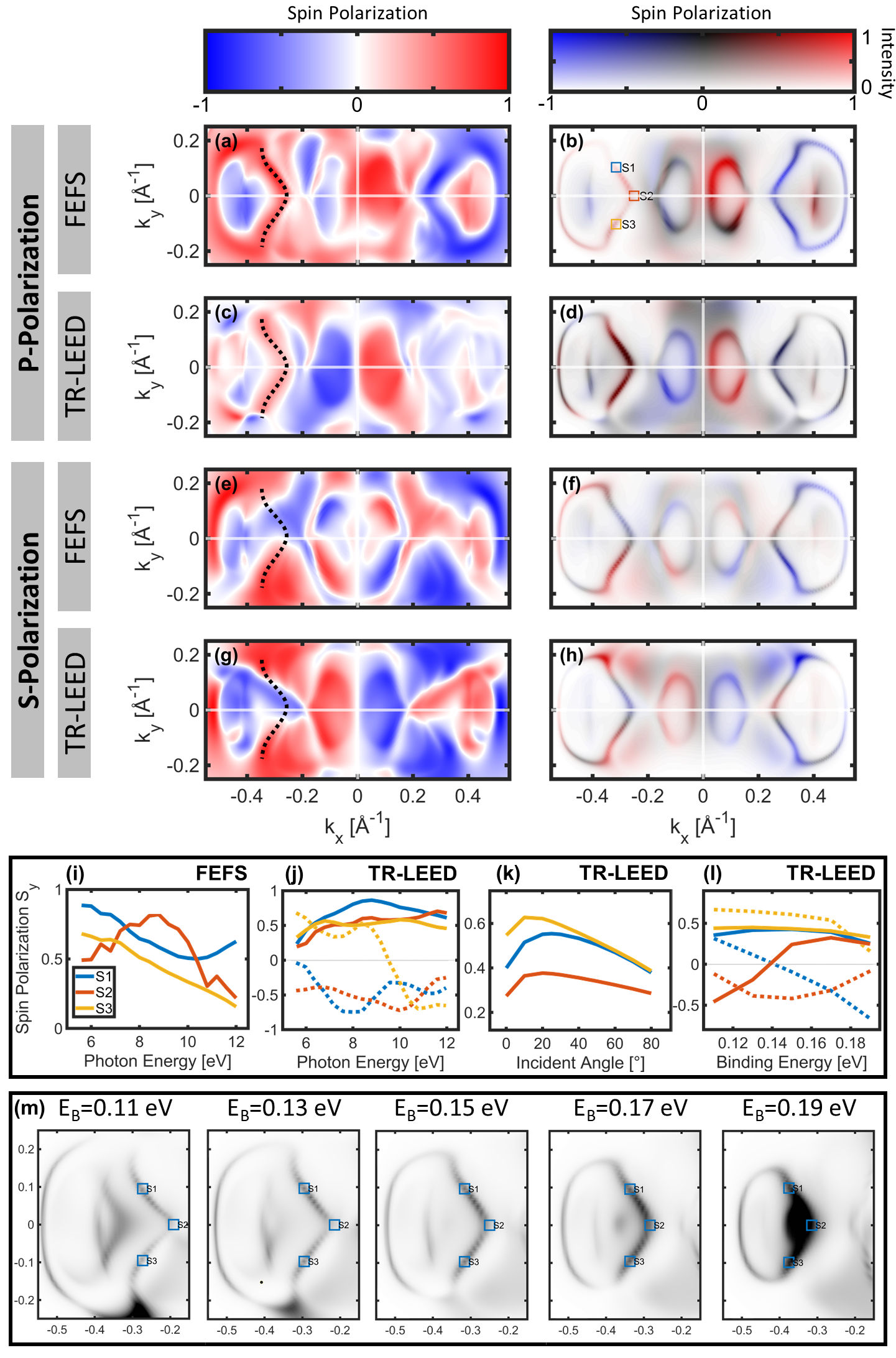}
    \caption{One-step-model calculation of spin polarization $S_y$ (a-d) $p$-polarized and (e-h) $s$-polarized light. (a), (c), (e) and (g) show pure spin polarization, while (b), (d), (f), and (h) are weighted by photoemission intensity. For (a-b) and (e-f) the FEFS was used, while for (c-d) and (g-h) the TR-LEED final state was used. (i-k) depict the magnitude of $S_y$ at 3 momenta $S1$-$S3$ indicated in (b), solid lines are for $p$-polarized light, and dashed lines in (j) and (l) are for $s$-polarized light. (i) and (j) show the dependence on photon energy for the FEFS and TR-LEED final states, respectively. (k) shows the dependence on the off-normal light incidence angle. (l) shows the dependence on binding energy for the momenta indicated in (m), where TR-LEED maps (with $p$-polarized light) at several binding energies are shown. For convenience the dashed lines in (a), (c), (e), and (g) indicate the location of the surface state.}
    \label{fig:onestep}
\end{figure}

Figure \ref{fig:onestep}(a-h) presents one-step model calculations of the spin-polarization from WTe$_2$ at our experimental parameters, and at the theoretical KKR binding energy E$_B = 0.15$ eV that best matches the experimental Fermi level (see Section SIII of the Supplemental Material for details). For both the $p$- and $s$-polarized light and for both the free-electron final state (FEFS) and time-reversed (TR) LEED final states, the $S_{fy}$ ARPES spin textures exhibit broken $\mathcal M_y$ symmetry, unlike in the initial state map in Fig. \ref{fig:domains}(c), but in qualitative agreement with the experiment. Considering the system of a sample together with an incident light, for $s$-polarized light the $\mathcal M_x$ symmetry of the WTe$_2$ is conserved, because in this case the $\mathbf E$ field of the light is along the $y$ axis, leading to $S_{fy}(k_{fx},k_{fy}) = -S_{fy}(-k_{fx},k_{fy})$. For the $p$-polarized light $\mathcal M_x$ is broken due to $\mathbf E$ being within the $xz$ plane.

Fig. \ref{fig:onestep} (i-m) theoretically explore the parameter space for $p$- and $s$-polarized light. While the asymmetry effect is firmly confirmed, for both polarizations it strongly depends on photon and binding energies. Furthermore, in contrast to experiment, at the experimental parameters its magnitude is predicted to be smaller for $p$-polarized light than for $s$-polarized light.



{\it Discussion.---} The origin of the spin-filtering effect can be understood within the tight-binding (TB) formalism taking into account the FEFS matrix element $w(k_{fx},k_{fy},k_{fy})=\langle e^{i\mathbf{k}_f \cdot \mathbf{r}}|\psi_i(k_{ix},k_{iy},\mathbf r)\rangle$ \cite{Moser2017,Zhu2013}, where $\psi_i$ is the initial surface state eigenfunction with an eigenvalue (binding energy) $E_B$. Assuming a system of $N$ orbitals $|j\rangle$ at positions $\mathbf r_j$ we can write $\psi_i(k_{ix},k_{iy},\mathbf r)=\sum_{j=1}^N C_j |j\rangle \delta(\mathbf r - \mathbf r_j)$.  The FEFS matrix element is essentially a Fourier transform and therefore different sites $\mathbf r_j$ will lead to phase shifts $e^{i \mathbf{k}_f\cdot \mathbf{r}_j}$. Since within the TB formalism only discrete sites $\mathbf r_j$ are considered, the matrix element will have a form of a sum $w(k_{fx},k_{fy},k_{fz})=\sum_{j=1}^N e^{i \mathbf{k}_f \cdot \mathbf{r}_j} C_j |j\rangle$.

Relating to squares $S1$ and $S3$ in Fig. \ref{fig:onestep}(b), for a periodic system, probing initial parallel momenta $\mathbf k_{iS1} = (k_{ix},k_{iy})$ and $\mathbf k_{iS3} =(k_{ix},-k_{iy})$ by ARPES requires measuring electrons emitted along $\mathbf k_{fS1} = (k_{ix},k_{iy},k_{fz})$ and $\mathbf k_{fS3} = (k_{ix},-k_{iy},k_{fz})$. For a particular photon energy $h\nu$, $k_{fz}$ can be determined through $|\mathbf k_f| = (1/\hbar)\sqrt{2m E_{kin}}$, where $E_{kin} = hv-W-E_B$ and $W$ is the work function, making the model $h\nu$-dependent.

Since atomic sites are positioned such that $\mathcal M_y$ is broken, this will lead to different $\mathbf r_j$-derived phase shifts in $\langle e^{i \mathbf{k}_{fS1} \cdot \mathbf{r}}| \psi_i(\mathbf k_{iS1},\mathbf r)\rangle$ and $\langle e^{i \mathbf{k}_{fS3} \cdot \mathbf{r}}|\psi_i(\mathbf k_{iS3},\mathbf r)\rangle$. One can show that in a generic case this leads to different spin polarizations for the two emission directions, $\mathbf k_{fS1}$ and $\mathbf k_{fS3}$, despite equal initial state polarizations $S_{iy}(\mathbf k_{fS1})$ and $S_{iy}(\mathbf k_{fS3})$, as discussed earlier.  A full derivation of this interference model for the minimal case of 2 orbitals on 2 different sites, that is lengthy but elementary, is presented in Section SIV of the Supplemental Material, while the complex orbital structure (Fig. \ref{fig:domains}(d)) forbids writing a realistic TB model of WTe$_2$. The existence of the asymmetry for $s$-polarized light (Supplemental Material Fig. S4) is in agreement with the interference model and points out that the effect is not related to breaking of the $\mathcal M_x$ mirror plane by the $p$-polarized light.

Fig.~\ref{fig:onestep}(i-m) illustrates sensitive dependence of the SARPES spin texture on various parameters. Importantly, the predicted effect at the experimental parameters ($p$-polarized light, $h\nu$ = 6 eV, light incidence angle $80^\circ$) is small, which calls for further improvement of the theoretical description that is critical in establishing whether WTe$_2$ realizes a Weyl type-II phase. These improvements can include further exploring correlation effects \cite{DiSante2017}, and structural changes \cite{Rossi2020,Xiao2020} and relaxations, which can lead to significant renormalization of the electronic structure and orbital character of the surface states.

In this spirit, we propose that establishing initial state spin textures experimentally shall involve iterative optimization of the agreement to the initial state and one-step model calculations, exploring the parameter space such as in Fig.~\ref{fig:onestep}(i-m). Experimentally this is currently challenging, but might be feasible using the new generation spin detectors \cite{Tusche2015,Schoenhense2015}.

Following the above arguments, we propose that spin asymmetries as observed here shall be present in SARPES from every surface that exhibits spin-momentum locked spin-polarized bands and lacks the mirror plane perpendicular to one of the lateral directions. Previously similar effects have been studied theoretically, but not directly discussed, for topological insulators \cite{Zhu2013} and Rashba systems \cite{Bentmann2017}.


The presented effect is different from the interference photoemission models for the generic orientation of the light polarization \cite{Yaji2017}, since these models do not take into account phase shifts in matrix element due to different locations of atomic sites. Spin filtering in photoemission through ultrathin ferromagentic layers has been discussed previously \cite{Henk2003}, however, the effect in WTe$_2$ is different since it does not involve time-reversal symmetry breaking in the initial state. Conversely, a filtering due to non-magnetic overlayers on magnetic substrates has also been studied \cite{Berdot2010}, which is again different from the effect discussed here, since it involves a modification of the initial band structure.


{\it Summary.---} Low crystal symmetry of WTe$_2$, which is responsible for its exotic transport properties, leads to emerging asymmetries in the SARPES spin textures. We have characterized spin textures in the photoelectron ensemble from WTe$_2$ surfaces excited by the 6 eV continuous wave laser light. Despite the overall high asymmetry, the spin textures of adjacent terraces are connected by the $\mathcal M_y$ mirror symmetry operation. Results have been obtained using a novel SARPES machine that enables 2D mapping of the spin textures with reduced instrumental asymmetries. The modulation of the photoelectron spin-polarization can be interpreted as the geometry-induced, light-polarization-independent spin filtering effect which can be modeled qualitatively within the free-electron final state photoemission model with the $\langle e^{i \mathbf{k}_f  \cdot \mathbf{r}} | \psi(\mathbf{k},\mathbf{r}) \rangle$ matrix element.

These results are the first demonstration of the surface spin texture of WTe$_2$ over the extended momentum range, and demonstrate a new aspect of a non-trivial connection between the initial state band structure properties and the photoelectron constant energy reciprocal space maps. Similar effects are expected in other quantum materials where a corresponding experimental geometry can be established. A more complete picture could be obtained by combining circular-dichroic and spin-texture ARPES maps which could address the orbital character of the participating states towards the identification of the transport-relevant Berry curvature hot spots. In this way, our results call for future research to establish a possible connection to the spin transport of hot electrons inside the crystal, through the interfaces, and through the surface barrier.



\begin{acknowledgments}

Authors would like to thank F. Freimuth, D. Nabok, S. Ghosh, J. Henk and Ph. R\"u\ss mann for fruitful discussions. Moreover, L. P. and T. H. acknowledge the support of Peter Baltzer (MB Scientific), Nicola Gatti (LEOS Solutions), and Matthias Escher (Focus GmbH). At different stages of this project the position of T. H. was funded by the Deutsche Forschungsgemeinschaft (DFG, German Research Foundation) under Germany's Excellence Strategy – Cluster of Excellence Matter and Light for Quantum Computing (ML4Q) EXC 2004/1 – 390534769 and by the DFG Priority Program SPP1666. J. M. and J. S. would like to thank the CEDAMNF (Grant No. CZ.02.1.01/0.0/0.0/15\_003/0000358) co-funded by the Ministry of Education, Youth and Sports of Czech Republic and the GACR Project No. 20-18725S for funding. J. S. and F. R. acknowledge financial support from the DFG through the W\"urzburg-Dresden Cluster of Excellence on Complexity and Topology in Quantum Matter -- \textit{ct.qmat} (EXC 2147, project-id 39085490). G. B. gratefully acknowledges the computing time granted through JARA-HPC on the supercomputer JURECA at Forschungszentrum J\"ulich.

\end{acknowledgments}



\begin{thebibliography}{37}%
\makeatletter
\providecommand \@ifxundefined [1]{%
 \@ifx{#1\undefined}
}%
\providecommand \@ifnum [1]{%
 \ifnum #1\expandafter \@firstoftwo
 \else \expandafter \@secondoftwo
 \fi
}%
\providecommand \@ifx [1]{%
 \ifx #1\expandafter \@firstoftwo
 \else \expandafter \@secondoftwo
 \fi
}%
\providecommand \natexlab [1]{#1}%
\providecommand \enquote  [1]{``#1''}%
\providecommand \bibnamefont  [1]{#1}%
\providecommand \bibfnamefont [1]{#1}%
\providecommand \citenamefont [1]{#1}%
\providecommand \href@noop [0]{\@secondoftwo}%
\providecommand \href [0]{\begingroup \@sanitize@url \@href}%
\providecommand \@href[1]{\@@startlink{#1}\@@href}%
\providecommand \@@href[1]{\endgroup#1\@@endlink}%
\providecommand \@sanitize@url [0]{\catcode `\\12\catcode `\$12\catcode
  `\&12\catcode `\#12\catcode `\^12\catcode `\_12\catcode `\%12\relax}%
\providecommand \@@startlink[1]{}%
\providecommand \@@endlink[0]{}%
\providecommand \url  [0]{\begingroup\@sanitize@url \@url }%
\providecommand \@url [1]{\endgroup\@href {#1}{\urlprefix }}%
\providecommand \urlprefix  [0]{URL }%
\providecommand \Eprint [0]{\href }%
\providecommand \doibase [0]{https://doi.org/}%
\providecommand \selectlanguage [0]{\@gobble}%
\providecommand \bibinfo  [0]{\@secondoftwo}%
\providecommand \bibfield  [0]{\@secondoftwo}%
\providecommand \translation [1]{[#1]}%
\providecommand \BibitemOpen [0]{}%
\providecommand \bibitemStop [0]{}%
\providecommand \bibitemNoStop [0]{.\EOS\space}%
\providecommand \EOS [0]{\spacefactor3000\relax}%
\providecommand \BibitemShut  [1]{\csname bibitem#1\endcsname}%
\let\auto@bib@innerbib\@empty
\bibitem [{\citenamefont {Ali}\ \emph {et~al.}(2014)\citenamefont {Ali},
  \citenamefont {Xiong}, \citenamefont {Flynn}, \citenamefont {Tao},
  \citenamefont {Gibson}, \citenamefont {Schoop}, \citenamefont {Liang},
  \citenamefont {Haldolaarachchige}, \citenamefont {Hirschberger},
  \citenamefont {Ong},\ and\ \citenamefont {Cava}}]{Ali2014}%
  \BibitemOpen
  \bibfield  {author} {\bibinfo {author} {\bibfnamefont {M.~N.}\ \bibnamefont
  {Ali}}, \bibinfo {author} {\bibfnamefont {J.}~\bibnamefont {Xiong}}, \bibinfo
  {author} {\bibfnamefont {S.}~\bibnamefont {Flynn}}, \bibinfo {author}
  {\bibfnamefont {J.}~\bibnamefont {Tao}}, \bibinfo {author} {\bibfnamefont
  {Q.~D.}\ \bibnamefont {Gibson}}, \bibinfo {author} {\bibfnamefont {L.~M.}\
  \bibnamefont {Schoop}}, \bibinfo {author} {\bibfnamefont {T.}~\bibnamefont
  {Liang}}, \bibinfo {author} {\bibfnamefont {N.}~\bibnamefont
  {Haldolaarachchige}}, \bibinfo {author} {\bibfnamefont {M.}~\bibnamefont
  {Hirschberger}}, \bibinfo {author} {\bibfnamefont {N.~P.}\ \bibnamefont
  {Ong}},\ and\ \bibinfo {author} {\bibfnamefont {R.~J.}\ \bibnamefont
  {Cava}},\ }\bibfield  {title} {\bibinfo {title} {Large, non-saturating
  magnetoresistance in {WTe}2},\ }\href {https://doi.org/10.1038/nature13763}
  {\bibfield  {journal} {\bibinfo  {journal} {Nature}\ }\textbf {\bibinfo
  {volume} {514}},\ \bibinfo {pages} {205} (\bibinfo {year}
  {2014})}\BibitemShut {NoStop}%
\bibitem [{\citenamefont {Soluyanov}\ \emph {et~al.}(2015)\citenamefont
  {Soluyanov}, \citenamefont {Gresch}, \citenamefont {Wang}, \citenamefont
  {Wu}, \citenamefont {Troyer}, \citenamefont {Dai},\ and\ \citenamefont
  {Bernevig}}]{Soluyanov2015}%
  \BibitemOpen
  \bibfield  {author} {\bibinfo {author} {\bibfnamefont {A.~A.}\ \bibnamefont
  {Soluyanov}}, \bibinfo {author} {\bibfnamefont {D.}~\bibnamefont {Gresch}},
  \bibinfo {author} {\bibfnamefont {Z.}~\bibnamefont {Wang}}, \bibinfo {author}
  {\bibfnamefont {Q.}~\bibnamefont {Wu}}, \bibinfo {author} {\bibfnamefont
  {M.}~\bibnamefont {Troyer}}, \bibinfo {author} {\bibfnamefont
  {X.}~\bibnamefont {Dai}},\ and\ \bibinfo {author} {\bibfnamefont {B.~A.}\
  \bibnamefont {Bernevig}},\ }\bibfield  {title} {\bibinfo {title} {Type-{II}
  weyl semimetals},\ }\href {https://doi.org/10.1038/nature15768} {\bibfield
  {journal} {\bibinfo  {journal} {Nature}\ }\textbf {\bibinfo {volume} {527}},\
  \bibinfo {pages} {495} (\bibinfo {year} {2015})}\BibitemShut {NoStop}%
\bibitem [{\citenamefont {R\"u\ss{}mann}\ \emph {et~al.}(2018)\citenamefont
  {R\"u\ss{}mann}, \citenamefont {Weber}, \citenamefont {Glott}, \citenamefont
  {Xu}, \citenamefont {Fanciulli}, \citenamefont {Muff}, \citenamefont
  {Magrez}, \citenamefont {Bugnon}, \citenamefont {Berger}, \citenamefont
  {Bode}, \citenamefont {Dil}, \citenamefont {Bl\"ugel}, \citenamefont
  {Mavropoulos},\ and\ \citenamefont {Sessi}}]{Ruessmann2018}%
  \BibitemOpen
  \bibfield  {author} {\bibinfo {author} {\bibfnamefont {P.}~\bibnamefont
  {R\"u\ss{}mann}}, \bibinfo {author} {\bibfnamefont {A.~P.}\ \bibnamefont
  {Weber}}, \bibinfo {author} {\bibfnamefont {F.}~\bibnamefont {Glott}},
  \bibinfo {author} {\bibfnamefont {N.}~\bibnamefont {Xu}}, \bibinfo {author}
  {\bibfnamefont {M.}~\bibnamefont {Fanciulli}}, \bibinfo {author}
  {\bibfnamefont {S.}~\bibnamefont {Muff}}, \bibinfo {author} {\bibfnamefont
  {A.}~\bibnamefont {Magrez}}, \bibinfo {author} {\bibfnamefont
  {P.}~\bibnamefont {Bugnon}}, \bibinfo {author} {\bibfnamefont
  {H.}~\bibnamefont {Berger}}, \bibinfo {author} {\bibfnamefont
  {M.}~\bibnamefont {Bode}}, \bibinfo {author} {\bibfnamefont {J.~H.}\
  \bibnamefont {Dil}}, \bibinfo {author} {\bibfnamefont {S.}~\bibnamefont
  {Bl\"ugel}}, \bibinfo {author} {\bibfnamefont {P.}~\bibnamefont
  {Mavropoulos}},\ and\ \bibinfo {author} {\bibfnamefont {P.}~\bibnamefont
  {Sessi}},\ }\bibfield  {title} {\bibinfo {title} {Universal scattering
  response across the type-ii weyl semimetal phase diagram},\ }\href
  {https://doi.org/10.1103/PhysRevB.97.075106} {\bibfield  {journal} {\bibinfo
  {journal} {Phys. Rev. B}\ }\textbf {\bibinfo {volume} {97}},\ \bibinfo
  {pages} {075106} (\bibinfo {year} {2018})}\BibitemShut {NoStop}%
\bibitem [{\citenamefont {Bruno}\ \emph {et~al.}(2016)\citenamefont {Bruno},
  \citenamefont {Tamai}, \citenamefont {Wu}, \citenamefont {Cucchi},
  \citenamefont {Barreteau}, \citenamefont {de~la Torre}, \citenamefont
  {McKeown~Walker}, \citenamefont {Ricc\`o}, \citenamefont {Wang},
  \citenamefont {Kim}, \citenamefont {Hoesch}, \citenamefont {Shi},
  \citenamefont {Plumb}, \citenamefont {Giannini}, \citenamefont {Soluyanov},\
  and\ \citenamefont {Baumberger}}]{Bruno2016}%
  \BibitemOpen
  \bibfield  {author} {\bibinfo {author} {\bibfnamefont {F.~Y.}\ \bibnamefont
  {Bruno}}, \bibinfo {author} {\bibfnamefont {A.}~\bibnamefont {Tamai}},
  \bibinfo {author} {\bibfnamefont {Q.~S.}\ \bibnamefont {Wu}}, \bibinfo
  {author} {\bibfnamefont {I.}~\bibnamefont {Cucchi}}, \bibinfo {author}
  {\bibfnamefont {C.}~\bibnamefont {Barreteau}}, \bibinfo {author}
  {\bibfnamefont {A.}~\bibnamefont {de~la Torre}}, \bibinfo {author}
  {\bibfnamefont {S.}~\bibnamefont {McKeown~Walker}}, \bibinfo {author}
  {\bibfnamefont {S.}~\bibnamefont {Ricc\`o}}, \bibinfo {author} {\bibfnamefont
  {Z.}~\bibnamefont {Wang}}, \bibinfo {author} {\bibfnamefont {T.~K.}\
  \bibnamefont {Kim}}, \bibinfo {author} {\bibfnamefont {M.}~\bibnamefont
  {Hoesch}}, \bibinfo {author} {\bibfnamefont {M.}~\bibnamefont {Shi}},
  \bibinfo {author} {\bibfnamefont {N.~C.}\ \bibnamefont {Plumb}}, \bibinfo
  {author} {\bibfnamefont {E.}~\bibnamefont {Giannini}}, \bibinfo {author}
  {\bibfnamefont {A.~A.}\ \bibnamefont {Soluyanov}},\ and\ \bibinfo {author}
  {\bibfnamefont {F.}~\bibnamefont {Baumberger}},\ }\bibfield  {title}
  {\bibinfo {title} {Observation of large topologically trivial fermi arcs in
  the candidate type-ii weyl semimetal $\mathrm{WT}{\mathrm{e}}_{2}$},\ }\href
  {https://doi.org/10.1103/PhysRevB.94.121112} {\bibfield  {journal} {\bibinfo
  {journal} {Phys. Rev. B}\ }\textbf {\bibinfo {volume} {94}},\ \bibinfo
  {pages} {121112} (\bibinfo {year} {2016})}\BibitemShut {NoStop}%
\bibitem [{\citenamefont {Tang}\ \emph {et~al.}(2017)\citenamefont {Tang},
  \citenamefont {Zhang}, \citenamefont {Wong}, \citenamefont {Pedramrazi},
  \citenamefont {Tsai}, \citenamefont {Jia}, \citenamefont {Moritz},
  \citenamefont {Claassen}, \citenamefont {Ryu}, \citenamefont {Kahn},
  \citenamefont {Jiang}, \citenamefont {Yan}, \citenamefont {Hashimoto},
  \citenamefont {Lu}, \citenamefont {Moore}, \citenamefont {Hwang},
  \citenamefont {Hwang}, \citenamefont {Hussain}, \citenamefont {Chen},
  \citenamefont {Ugeda}, \citenamefont {Liu}, \citenamefont {Xie},
  \citenamefont {Devereaux}, \citenamefont {Crommie}, \citenamefont {Mo},\ and\
  \citenamefont {Shen}}]{Tang2017}%
  \BibitemOpen
  \bibfield  {author} {\bibinfo {author} {\bibfnamefont {S.}~\bibnamefont
  {Tang}}, \bibinfo {author} {\bibfnamefont {C.}~\bibnamefont {Zhang}},
  \bibinfo {author} {\bibfnamefont {D.}~\bibnamefont {Wong}}, \bibinfo {author}
  {\bibfnamefont {Z.}~\bibnamefont {Pedramrazi}}, \bibinfo {author}
  {\bibfnamefont {H.-Z.}\ \bibnamefont {Tsai}}, \bibinfo {author}
  {\bibfnamefont {C.}~\bibnamefont {Jia}}, \bibinfo {author} {\bibfnamefont
  {B.}~\bibnamefont {Moritz}}, \bibinfo {author} {\bibfnamefont
  {M.}~\bibnamefont {Claassen}}, \bibinfo {author} {\bibfnamefont
  {H.}~\bibnamefont {Ryu}}, \bibinfo {author} {\bibfnamefont {S.}~\bibnamefont
  {Kahn}}, \bibinfo {author} {\bibfnamefont {J.}~\bibnamefont {Jiang}},
  \bibinfo {author} {\bibfnamefont {H.}~\bibnamefont {Yan}}, \bibinfo {author}
  {\bibfnamefont {M.}~\bibnamefont {Hashimoto}}, \bibinfo {author}
  {\bibfnamefont {D.}~\bibnamefont {Lu}}, \bibinfo {author} {\bibfnamefont
  {R.~G.}\ \bibnamefont {Moore}}, \bibinfo {author} {\bibfnamefont {C.-C.}\
  \bibnamefont {Hwang}}, \bibinfo {author} {\bibfnamefont {C.}~\bibnamefont
  {Hwang}}, \bibinfo {author} {\bibfnamefont {Z.}~\bibnamefont {Hussain}},
  \bibinfo {author} {\bibfnamefont {Y.}~\bibnamefont {Chen}}, \bibinfo {author}
  {\bibfnamefont {M.~M.}\ \bibnamefont {Ugeda}}, \bibinfo {author}
  {\bibfnamefont {Z.}~\bibnamefont {Liu}}, \bibinfo {author} {\bibfnamefont
  {X.}~\bibnamefont {Xie}}, \bibinfo {author} {\bibfnamefont {T.~P.}\
  \bibnamefont {Devereaux}}, \bibinfo {author} {\bibfnamefont {M.~F.}\
  \bibnamefont {Crommie}}, \bibinfo {author} {\bibfnamefont {S.-K.}\
  \bibnamefont {Mo}},\ and\ \bibinfo {author} {\bibfnamefont {Z.-X.}\
  \bibnamefont {Shen}},\ }\bibfield  {title} {\bibinfo {title} {Quantum spin
  hall state in monolayer 1t{\textquotesingle}-{WTe}2},\ }\href
  {https://doi.org/10.1038/nphys4174} {\bibfield  {journal} {\bibinfo
  {journal} {Nature Physics}\ }\textbf {\bibinfo {volume} {13}},\ \bibinfo
  {pages} {683} (\bibinfo {year} {2017})}\BibitemShut {NoStop}%
\bibitem [{\citenamefont {Fatemi}\ \emph {et~al.}(2018)\citenamefont {Fatemi},
  \citenamefont {Wu}, \citenamefont {Cao}, \citenamefont {Bretheau},
  \citenamefont {Gibson}, \citenamefont {Watanabe}, \citenamefont {Taniguchi},
  \citenamefont {Cava},\ and\ \citenamefont {Jarillo-Herrero}}]{Fatemi2018}%
  \BibitemOpen
  \bibfield  {author} {\bibinfo {author} {\bibfnamefont {V.}~\bibnamefont
  {Fatemi}}, \bibinfo {author} {\bibfnamefont {S.}~\bibnamefont {Wu}}, \bibinfo
  {author} {\bibfnamefont {Y.}~\bibnamefont {Cao}}, \bibinfo {author}
  {\bibfnamefont {L.}~\bibnamefont {Bretheau}}, \bibinfo {author}
  {\bibfnamefont {Q.~D.}\ \bibnamefont {Gibson}}, \bibinfo {author}
  {\bibfnamefont {K.}~\bibnamefont {Watanabe}}, \bibinfo {author}
  {\bibfnamefont {T.}~\bibnamefont {Taniguchi}}, \bibinfo {author}
  {\bibfnamefont {R.~J.}\ \bibnamefont {Cava}},\ and\ \bibinfo {author}
  {\bibfnamefont {P.}~\bibnamefont {Jarillo-Herrero}},\ }\bibfield  {title}
  {\bibinfo {title} {Electrically tunable low-density superconductivity in a
  monolayer topological insulator},\ }\href
  {https://doi.org/10.1126/science.aar4642} {\bibfield  {journal} {\bibinfo
  {journal} {Science}\ }\textbf {\bibinfo {volume} {362}},\ \bibinfo {pages}
  {926} (\bibinfo {year} {2018})},\ \Eprint
  {https://arxiv.org/abs/https://science.sciencemag.org/content/362/6417/926.full.pdf}
  {https://science.sciencemag.org/content/362/6417/926.full.pdf} \BibitemShut
  {NoStop}%
\bibitem [{\citenamefont {Sajadi}\ \emph {et~al.}(2018)\citenamefont {Sajadi},
  \citenamefont {Palomaki}, \citenamefont {Fei}, \citenamefont {Zhao},
  \citenamefont {Bement}, \citenamefont {Olsen}, \citenamefont {Luescher},
  \citenamefont {Xu}, \citenamefont {Folk},\ and\ \citenamefont
  {Cobden}}]{Sajadi2018}%
  \BibitemOpen
  \bibfield  {author} {\bibinfo {author} {\bibfnamefont {E.}~\bibnamefont
  {Sajadi}}, \bibinfo {author} {\bibfnamefont {T.}~\bibnamefont {Palomaki}},
  \bibinfo {author} {\bibfnamefont {Z.}~\bibnamefont {Fei}}, \bibinfo {author}
  {\bibfnamefont {W.}~\bibnamefont {Zhao}}, \bibinfo {author} {\bibfnamefont
  {P.}~\bibnamefont {Bement}}, \bibinfo {author} {\bibfnamefont
  {C.}~\bibnamefont {Olsen}}, \bibinfo {author} {\bibfnamefont
  {S.}~\bibnamefont {Luescher}}, \bibinfo {author} {\bibfnamefont
  {X.}~\bibnamefont {Xu}}, \bibinfo {author} {\bibfnamefont {J.~A.}\
  \bibnamefont {Folk}},\ and\ \bibinfo {author} {\bibfnamefont {D.~H.}\
  \bibnamefont {Cobden}},\ }\bibfield  {title} {\bibinfo {title} {Gate-induced
  superconductivity in a monolayer topological insulator},\ }\href
  {https://doi.org/10.1126/science.aar4426} {\bibfield  {journal} {\bibinfo
  {journal} {Science}\ }\textbf {\bibinfo {volume} {362}},\ \bibinfo {pages}
  {922} (\bibinfo {year} {2018})},\ \Eprint
  {https://arxiv.org/abs/https://science.sciencemag.org/content/362/6417/922.full.pdf}
  {https://science.sciencemag.org/content/362/6417/922.full.pdf} \BibitemShut
  {NoStop}%
\bibitem [{\citenamefont {Sodemann}\ and\ \citenamefont
  {Fu}(2015)}]{Sodemann2015}%
  \BibitemOpen
  \bibfield  {author} {\bibinfo {author} {\bibfnamefont {I.}~\bibnamefont
  {Sodemann}}\ and\ \bibinfo {author} {\bibfnamefont {L.}~\bibnamefont {Fu}},\
  }\bibfield  {title} {\bibinfo {title} {Quantum nonlinear hall effect induced
  by berry curvature dipole in time-reversal invariant materials},\ }\href
  {https://doi.org/10.1103/PhysRevLett.115.216806} {\bibfield  {journal}
  {\bibinfo  {journal} {Phys. Rev. Lett.}\ }\textbf {\bibinfo {volume} {115}},\
  \bibinfo {pages} {216806} (\bibinfo {year} {2015})}\BibitemShut {NoStop}%
\bibitem [{\citenamefont {Ma}\ \emph {et~al.}(2019)\citenamefont {Ma},
  \citenamefont {Xu}, \citenamefont {Shen}, \citenamefont {MacNeill},
  \citenamefont {Fatemi}, \citenamefont {Chang}, \citenamefont {Valdivia},
  \citenamefont {Wu}, \citenamefont {Du}, \citenamefont {Hsu}, \citenamefont
  {Fang}, \citenamefont {Gibson}, \citenamefont {Watanabe}, \citenamefont
  {Taniguchi}, \citenamefont {Cava}, \citenamefont {Kaxiras}, \citenamefont
  {Lu}, \citenamefont {Lin}, \citenamefont {Fu}, \citenamefont {Gedik},\ and\
  \citenamefont {Jarillo-Herrero}}]{Ma2019}%
  \BibitemOpen
  \bibfield  {author} {\bibinfo {author} {\bibfnamefont {Q.}~\bibnamefont
  {Ma}}, \bibinfo {author} {\bibfnamefont {S.-Y.}\ \bibnamefont {Xu}}, \bibinfo
  {author} {\bibfnamefont {H.}~\bibnamefont {Shen}}, \bibinfo {author}
  {\bibfnamefont {D.}~\bibnamefont {MacNeill}}, \bibinfo {author}
  {\bibfnamefont {V.}~\bibnamefont {Fatemi}}, \bibinfo {author} {\bibfnamefont
  {T.-R.}\ \bibnamefont {Chang}}, \bibinfo {author} {\bibfnamefont {A.~M.~M.}\
  \bibnamefont {Valdivia}}, \bibinfo {author} {\bibfnamefont {S.}~\bibnamefont
  {Wu}}, \bibinfo {author} {\bibfnamefont {Z.}~\bibnamefont {Du}}, \bibinfo
  {author} {\bibfnamefont {C.-H.}\ \bibnamefont {Hsu}}, \bibinfo {author}
  {\bibfnamefont {S.}~\bibnamefont {Fang}}, \bibinfo {author} {\bibfnamefont
  {Q.~D.}\ \bibnamefont {Gibson}}, \bibinfo {author} {\bibfnamefont
  {K.}~\bibnamefont {Watanabe}}, \bibinfo {author} {\bibfnamefont
  {T.}~\bibnamefont {Taniguchi}}, \bibinfo {author} {\bibfnamefont {R.~J.}\
  \bibnamefont {Cava}}, \bibinfo {author} {\bibfnamefont {E.}~\bibnamefont
  {Kaxiras}}, \bibinfo {author} {\bibfnamefont {H.-Z.}\ \bibnamefont {Lu}},
  \bibinfo {author} {\bibfnamefont {H.}~\bibnamefont {Lin}}, \bibinfo {author}
  {\bibfnamefont {L.}~\bibnamefont {Fu}}, \bibinfo {author} {\bibfnamefont
  {N.}~\bibnamefont {Gedik}},\ and\ \bibinfo {author} {\bibfnamefont
  {P.}~\bibnamefont {Jarillo-Herrero}},\ }\bibfield  {title} {\bibinfo {title}
  {Observation of the nonlinear hall effect under time-reversal-symmetric
  conditions},\ }\href {https://doi.org/10.1038/s41586-018-0807-6} {\bibfield
  {journal} {\bibinfo  {journal} {Nature}\ }\textbf {\bibinfo {volume} {565}},\
  \bibinfo {pages} {337} (\bibinfo {year} {2019})}\BibitemShut {NoStop}%
\bibitem [{\citenamefont {Kang}\ \emph {et~al.}(2019)\citenamefont {Kang},
  \citenamefont {Li}, \citenamefont {Sohn}, \citenamefont {Shan},\ and\
  \citenamefont {Mak}}]{Kang2019}%
  \BibitemOpen
  \bibfield  {author} {\bibinfo {author} {\bibfnamefont {K.}~\bibnamefont
  {Kang}}, \bibinfo {author} {\bibfnamefont {T.}~\bibnamefont {Li}}, \bibinfo
  {author} {\bibfnamefont {E.}~\bibnamefont {Sohn}}, \bibinfo {author}
  {\bibfnamefont {J.}~\bibnamefont {Shan}},\ and\ \bibinfo {author}
  {\bibfnamefont {K.~F.}\ \bibnamefont {Mak}},\ }\bibfield  {title} {\bibinfo
  {title} {Nonlinear anomalous hall effect in few-layer {WTe}2},\ }\href
  {https://doi.org/10.1038/s41563-019-0294-7} {\bibfield  {journal} {\bibinfo
  {journal} {Nature Materials}\ }\textbf {\bibinfo {volume} {18}},\ \bibinfo
  {pages} {324} (\bibinfo {year} {2019})}\BibitemShut {NoStop}%
\bibitem [{\citenamefont {Du}\ \emph {et~al.}(2021)\citenamefont {Du},
  \citenamefont {Lu},\ and\ \citenamefont {Xie}}]{Du2021}%
  \BibitemOpen
  \bibfield  {author} {\bibinfo {author} {\bibfnamefont {Z.~Z.}\ \bibnamefont
  {Du}}, \bibinfo {author} {\bibfnamefont {H.-Z.}\ \bibnamefont {Lu}},\ and\
  \bibinfo {author} {\bibfnamefont {X.~C.}\ \bibnamefont {Xie}},\ }\bibfield
  {title} {\bibinfo {title} {Nonlinear hall effects},\ }\bibfield  {journal}
  {\bibinfo  {journal} {Nature Reviews Physics}\ }\href
  {https://doi.org/10.1038/s42254-021-00359-6} {10.1038/s42254-021-00359-6}
  (\bibinfo {year} {2021})\BibitemShut {NoStop}%
\bibitem [{\citenamefont {Pletikosi\ifmmode~\acute{c}\else \'{c}\fi{}}\ \emph
  {et~al.}(2014)\citenamefont {Pletikosi\ifmmode~\acute{c}\else \'{c}\fi{}},
  \citenamefont {Ali}, \citenamefont {Fedorov}, \citenamefont {Cava},\ and\
  \citenamefont {Valla}}]{Pletikosic2014}%
  \BibitemOpen
  \bibfield  {author} {\bibinfo {author} {\bibfnamefont {I.}~\bibnamefont
  {Pletikosi\ifmmode~\acute{c}\else \'{c}\fi{}}}, \bibinfo {author}
  {\bibfnamefont {M.~N.}\ \bibnamefont {Ali}}, \bibinfo {author} {\bibfnamefont
  {A.~V.}\ \bibnamefont {Fedorov}}, \bibinfo {author} {\bibfnamefont {R.~J.}\
  \bibnamefont {Cava}},\ and\ \bibinfo {author} {\bibfnamefont
  {T.}~\bibnamefont {Valla}},\ }\bibfield  {title} {\bibinfo {title}
  {Electronic structure basis for the extraordinary magnetoresistance in
  ${\mathrm{wte}}_{2}$},\ }\href
  {https://doi.org/10.1103/PhysRevLett.113.216601} {\bibfield  {journal}
  {\bibinfo  {journal} {Phys. Rev. Lett.}\ }\textbf {\bibinfo {volume} {113}},\
  \bibinfo {pages} {216601} (\bibinfo {year} {2014})}\BibitemShut {NoStop}%
\bibitem [{\citenamefont {Wu}\ \emph {et~al.}(2015)\citenamefont {Wu},
  \citenamefont {Jo}, \citenamefont {Ochi}, \citenamefont {Huang},
  \citenamefont {Mou}, \citenamefont {Bud'ko}, \citenamefont {Canfield},
  \citenamefont {Trivedi}, \citenamefont {Arita},\ and\ \citenamefont
  {Kaminski}}]{Wu2015}%
  \BibitemOpen
  \bibfield  {author} {\bibinfo {author} {\bibfnamefont {Y.}~\bibnamefont
  {Wu}}, \bibinfo {author} {\bibfnamefont {N.~H.}\ \bibnamefont {Jo}}, \bibinfo
  {author} {\bibfnamefont {M.}~\bibnamefont {Ochi}}, \bibinfo {author}
  {\bibfnamefont {L.}~\bibnamefont {Huang}}, \bibinfo {author} {\bibfnamefont
  {D.}~\bibnamefont {Mou}}, \bibinfo {author} {\bibfnamefont {S.~L.}\
  \bibnamefont {Bud'ko}}, \bibinfo {author} {\bibfnamefont {P.~C.}\
  \bibnamefont {Canfield}}, \bibinfo {author} {\bibfnamefont {N.}~\bibnamefont
  {Trivedi}}, \bibinfo {author} {\bibfnamefont {R.}~\bibnamefont {Arita}},\
  and\ \bibinfo {author} {\bibfnamefont {A.}~\bibnamefont {Kaminski}},\
  }\bibfield  {title} {\bibinfo {title} {Temperature-induced lifshitz
  transition in ${\mathrm{wte}}_{2}$},\ }\href
  {https://doi.org/10.1103/PhysRevLett.115.166602} {\bibfield  {journal}
  {\bibinfo  {journal} {Phys. Rev. Lett.}\ }\textbf {\bibinfo {volume} {115}},\
  \bibinfo {pages} {166602} (\bibinfo {year} {2015})}\BibitemShut {NoStop}%
\bibitem [{\citenamefont {Di~Sante}\ \emph {et~al.}(2017)\citenamefont
  {Di~Sante}, \citenamefont {Das}, \citenamefont {Bigi}, \citenamefont
  {Erg\"onenc}, \citenamefont {G\"urtler}, \citenamefont {Krieger},
  \citenamefont {Schmitt}, \citenamefont {Ali}, \citenamefont {Rossi},
  \citenamefont {Thomale}, \citenamefont {Franchini}, \citenamefont {Picozzi},
  \citenamefont {Fujii}, \citenamefont {Strocov}, \citenamefont {Sangiovanni},
  \citenamefont {Vobornik}, \citenamefont {Cava},\ and\ \citenamefont
  {Panaccione}}]{DiSante2017}%
  \BibitemOpen
  \bibfield  {author} {\bibinfo {author} {\bibfnamefont {D.}~\bibnamefont
  {Di~Sante}}, \bibinfo {author} {\bibfnamefont {P.~K.}\ \bibnamefont {Das}},
  \bibinfo {author} {\bibfnamefont {C.}~\bibnamefont {Bigi}}, \bibinfo {author}
  {\bibfnamefont {Z.}~\bibnamefont {Erg\"onenc}}, \bibinfo {author}
  {\bibfnamefont {N.}~\bibnamefont {G\"urtler}}, \bibinfo {author}
  {\bibfnamefont {J.~A.}\ \bibnamefont {Krieger}}, \bibinfo {author}
  {\bibfnamefont {T.}~\bibnamefont {Schmitt}}, \bibinfo {author} {\bibfnamefont
  {M.~N.}\ \bibnamefont {Ali}}, \bibinfo {author} {\bibfnamefont
  {G.}~\bibnamefont {Rossi}}, \bibinfo {author} {\bibfnamefont
  {R.}~\bibnamefont {Thomale}}, \bibinfo {author} {\bibfnamefont
  {C.}~\bibnamefont {Franchini}}, \bibinfo {author} {\bibfnamefont
  {S.}~\bibnamefont {Picozzi}}, \bibinfo {author} {\bibfnamefont
  {J.}~\bibnamefont {Fujii}}, \bibinfo {author} {\bibfnamefont {V.~N.}\
  \bibnamefont {Strocov}}, \bibinfo {author} {\bibfnamefont {G.}~\bibnamefont
  {Sangiovanni}}, \bibinfo {author} {\bibfnamefont {I.}~\bibnamefont
  {Vobornik}}, \bibinfo {author} {\bibfnamefont {R.~J.}\ \bibnamefont {Cava}},\
  and\ \bibinfo {author} {\bibfnamefont {G.}~\bibnamefont {Panaccione}},\
  }\bibfield  {title} {\bibinfo {title} {Three-dimensional electronic structure
  of the type-ii weyl semimetal ${\mathrm{wte}}_{2}$},\ }\href
  {https://doi.org/10.1103/PhysRevLett.119.026403} {\bibfield  {journal}
  {\bibinfo  {journal} {Phys. Rev. Lett.}\ }\textbf {\bibinfo {volume} {119}},\
  \bibinfo {pages} {026403} (\bibinfo {year} {2017})}\BibitemShut {NoStop}%
\bibitem [{\citenamefont {Pippard}(1989)}]{Pippard1989}%
  \BibitemOpen
  \bibfield  {author} {\bibinfo {author} {\bibfnamefont {A.~B.}\ \bibnamefont
  {Pippard}},\ }\href@noop {} {\emph {\bibinfo {title} {Magnetoresistance in
  metals}}}\ (\bibinfo  {publisher} {Cambridge University Press},\ \bibinfo
  {address} {Cambridge England New York},\ \bibinfo {year} {1989})\BibitemShut
  {NoStop}%
\bibitem [{\citenamefont {Wang}\ \emph {et~al.}(2017)\citenamefont {Wang},
  \citenamefont {Zhang}, \citenamefont {Huang}, \citenamefont {Liu},
  \citenamefont {Liang}, \citenamefont {Zhang}, \citenamefont {Shen},
  \citenamefont {Liu}, \citenamefont {Hu}, \citenamefont {Ding}, \citenamefont
  {Liu}, \citenamefont {Hu}, \citenamefont {He}, \citenamefont {Zhao},
  \citenamefont {Yu}, \citenamefont {Hu}, \citenamefont {Wei}, \citenamefont
  {Mao}, \citenamefont {Shi}, \citenamefont {Jia}, \citenamefont {Zhang},
  \citenamefont {Zhang}, \citenamefont {Yang}, \citenamefont {Wang},
  \citenamefont {Peng}, \citenamefont {Xu}, \citenamefont {Chen},\ and\
  \citenamefont {Zhou}}]{Wang2017}%
  \BibitemOpen
  \bibfield  {author} {\bibinfo {author} {\bibfnamefont {C.-L.}\ \bibnamefont
  {Wang}}, \bibinfo {author} {\bibfnamefont {Y.}~\bibnamefont {Zhang}},
  \bibinfo {author} {\bibfnamefont {J.-W.}\ \bibnamefont {Huang}}, \bibinfo
  {author} {\bibfnamefont {G.-D.}\ \bibnamefont {Liu}}, \bibinfo {author}
  {\bibfnamefont {A.-J.}\ \bibnamefont {Liang}}, \bibinfo {author}
  {\bibfnamefont {Y.-X.}\ \bibnamefont {Zhang}}, \bibinfo {author}
  {\bibfnamefont {B.}~\bibnamefont {Shen}}, \bibinfo {author} {\bibfnamefont
  {J.}~\bibnamefont {Liu}}, \bibinfo {author} {\bibfnamefont {C.}~\bibnamefont
  {Hu}}, \bibinfo {author} {\bibfnamefont {Y.}~\bibnamefont {Ding}}, \bibinfo
  {author} {\bibfnamefont {D.-F.}\ \bibnamefont {Liu}}, \bibinfo {author}
  {\bibfnamefont {Y.}~\bibnamefont {Hu}}, \bibinfo {author} {\bibfnamefont
  {S.-L.}\ \bibnamefont {He}}, \bibinfo {author} {\bibfnamefont
  {L.}~\bibnamefont {Zhao}}, \bibinfo {author} {\bibfnamefont {L.}~\bibnamefont
  {Yu}}, \bibinfo {author} {\bibfnamefont {J.}~\bibnamefont {Hu}}, \bibinfo
  {author} {\bibfnamefont {J.}~\bibnamefont {Wei}}, \bibinfo {author}
  {\bibfnamefont {Z.-Q.}\ \bibnamefont {Mao}}, \bibinfo {author} {\bibfnamefont
  {Y.-G.}\ \bibnamefont {Shi}}, \bibinfo {author} {\bibfnamefont {X.-W.}\
  \bibnamefont {Jia}}, \bibinfo {author} {\bibfnamefont {F.-F.}\ \bibnamefont
  {Zhang}}, \bibinfo {author} {\bibfnamefont {S.-J.}\ \bibnamefont {Zhang}},
  \bibinfo {author} {\bibfnamefont {F.}~\bibnamefont {Yang}}, \bibinfo {author}
  {\bibfnamefont {Z.-M.}\ \bibnamefont {Wang}}, \bibinfo {author}
  {\bibfnamefont {Q.-J.}\ \bibnamefont {Peng}}, \bibinfo {author}
  {\bibfnamefont {Z.-Y.}\ \bibnamefont {Xu}}, \bibinfo {author} {\bibfnamefont
  {C.-T.}\ \bibnamefont {Chen}},\ and\ \bibinfo {author} {\bibfnamefont
  {X.-J.}\ \bibnamefont {Zhou}},\ }\bibfield  {title} {\bibinfo {title}
  {Evidence of electron-hole imbalance in {WTe}$_2$ from high-resolution
  angle-resolved photoemission spectroscopy},\ }\href
  {https://doi.org/10.1088/0256-307x/34/9/097305} {\bibfield  {journal}
  {\bibinfo  {journal} {Chinese Physics Letters}\ }\textbf {\bibinfo {volume}
  {34}},\ \bibinfo {pages} {097305} (\bibinfo {year} {2017})}\BibitemShut
  {NoStop}%
\bibitem [{\citenamefont {Wu}\ \emph {et~al.}(2017)\citenamefont {Wu},
  \citenamefont {Jo}, \citenamefont {Mou}, \citenamefont {Huang}, \citenamefont
  {Bud'ko}, \citenamefont {Canfield},\ and\ \citenamefont {Kaminski}}]{Wu2017}%
  \BibitemOpen
  \bibfield  {author} {\bibinfo {author} {\bibfnamefont {Y.}~\bibnamefont
  {Wu}}, \bibinfo {author} {\bibfnamefont {N.~H.}\ \bibnamefont {Jo}}, \bibinfo
  {author} {\bibfnamefont {D.}~\bibnamefont {Mou}}, \bibinfo {author}
  {\bibfnamefont {L.}~\bibnamefont {Huang}}, \bibinfo {author} {\bibfnamefont
  {S.~L.}\ \bibnamefont {Bud'ko}}, \bibinfo {author} {\bibfnamefont {P.~C.}\
  \bibnamefont {Canfield}},\ and\ \bibinfo {author} {\bibfnamefont
  {A.}~\bibnamefont {Kaminski}},\ }\bibfield  {title} {\bibinfo {title}
  {Three-dimensionality of the bulk electronic structure in
  ${\mathrm{wte}}_{2}$},\ }\href {https://doi.org/10.1103/PhysRevB.95.195138}
  {\bibfield  {journal} {\bibinfo  {journal} {Phys. Rev. B}\ }\textbf {\bibinfo
  {volume} {95}},\ \bibinfo {pages} {195138} (\bibinfo {year}
  {2017})}\BibitemShut {NoStop}%
\bibitem [{\citenamefont {Rossi}\ \emph {et~al.}(2020)\citenamefont {Rossi},
  \citenamefont {Resta}, \citenamefont {Lee}, \citenamefont {Redwing},
  \citenamefont {Jozwiak}, \citenamefont {Bostwick}, \citenamefont {Rotenberg},
  \citenamefont {Savrasov},\ and\ \citenamefont {Vishik}}]{Rossi2020}%
  \BibitemOpen
  \bibfield  {author} {\bibinfo {author} {\bibfnamefont {A.}~\bibnamefont
  {Rossi}}, \bibinfo {author} {\bibfnamefont {G.}~\bibnamefont {Resta}},
  \bibinfo {author} {\bibfnamefont {S.~H.}\ \bibnamefont {Lee}}, \bibinfo
  {author} {\bibfnamefont {R.~D.}\ \bibnamefont {Redwing}}, \bibinfo {author}
  {\bibfnamefont {C.}~\bibnamefont {Jozwiak}}, \bibinfo {author} {\bibfnamefont
  {A.}~\bibnamefont {Bostwick}}, \bibinfo {author} {\bibfnamefont
  {E.}~\bibnamefont {Rotenberg}}, \bibinfo {author} {\bibfnamefont {S.~Y.}\
  \bibnamefont {Savrasov}},\ and\ \bibinfo {author} {\bibfnamefont {I.~M.}\
  \bibnamefont {Vishik}},\ }\bibfield  {title} {\bibinfo {title} {Two phase
  transitions driven by surface electron doping in ${\mathrm{wte}}_{2}$},\
  }\href {https://doi.org/10.1103/PhysRevB.102.121110} {\bibfield  {journal}
  {\bibinfo  {journal} {Phys. Rev. B}\ }\textbf {\bibinfo {volume} {102}},\
  \bibinfo {pages} {121110} (\bibinfo {year} {2020})}\BibitemShut {NoStop}%
\bibitem [{\citenamefont {Feng}\ \emph {et~al.}(2016)\citenamefont {Feng},
  \citenamefont {Chan}, \citenamefont {Feng}, \citenamefont {Liu},
  \citenamefont {Chou}, \citenamefont {Kuroda}, \citenamefont {Yaji},
  \citenamefont {Harasawa}, \citenamefont {Moras}, \citenamefont {Barinov},
  \citenamefont {Malaeb}, \citenamefont {Bareille}, \citenamefont {Kondo},
  \citenamefont {Shin}, \citenamefont {Komori}, \citenamefont {Chiang},
  \citenamefont {Shi},\ and\ \citenamefont {Matsuda}}]{Feng2016}%
  \BibitemOpen
  \bibfield  {author} {\bibinfo {author} {\bibfnamefont {B.}~\bibnamefont
  {Feng}}, \bibinfo {author} {\bibfnamefont {Y.-H.}\ \bibnamefont {Chan}},
  \bibinfo {author} {\bibfnamefont {Y.}~\bibnamefont {Feng}}, \bibinfo {author}
  {\bibfnamefont {R.-Y.}\ \bibnamefont {Liu}}, \bibinfo {author} {\bibfnamefont
  {M.-Y.}\ \bibnamefont {Chou}}, \bibinfo {author} {\bibfnamefont
  {K.}~\bibnamefont {Kuroda}}, \bibinfo {author} {\bibfnamefont
  {K.}~\bibnamefont {Yaji}}, \bibinfo {author} {\bibfnamefont {A.}~\bibnamefont
  {Harasawa}}, \bibinfo {author} {\bibfnamefont {P.}~\bibnamefont {Moras}},
  \bibinfo {author} {\bibfnamefont {A.}~\bibnamefont {Barinov}}, \bibinfo
  {author} {\bibfnamefont {W.}~\bibnamefont {Malaeb}}, \bibinfo {author}
  {\bibfnamefont {C.}~\bibnamefont {Bareille}}, \bibinfo {author}
  {\bibfnamefont {T.}~\bibnamefont {Kondo}}, \bibinfo {author} {\bibfnamefont
  {S.}~\bibnamefont {Shin}}, \bibinfo {author} {\bibfnamefont {F.}~\bibnamefont
  {Komori}}, \bibinfo {author} {\bibfnamefont {T.-C.}\ \bibnamefont {Chiang}},
  \bibinfo {author} {\bibfnamefont {Y.}~\bibnamefont {Shi}},\ and\ \bibinfo
  {author} {\bibfnamefont {I.}~\bibnamefont {Matsuda}},\ }\bibfield  {title}
  {\bibinfo {title} {Spin texture in type-ii weyl semimetal
  ${\mathrm{wte}}_{2}$},\ }\href {https://doi.org/10.1103/PhysRevB.94.195134}
  {\bibfield  {journal} {\bibinfo  {journal} {Phys. Rev. B}\ }\textbf {\bibinfo
  {volume} {94}},\ \bibinfo {pages} {195134} (\bibinfo {year}
  {2016})}\BibitemShut {NoStop}%
\bibitem [{\citenamefont {Fanciulli}\ \emph {et~al.}(2020)\citenamefont
  {Fanciulli}, \citenamefont {Schusser}, \citenamefont {Lee}, \citenamefont
  {Youbi}, \citenamefont {Heckmann}, \citenamefont {Richter}, \citenamefont
  {Cacho}, \citenamefont {Spezzani}, \citenamefont {Bresteau}, \citenamefont
  {Hergott}, \citenamefont {D'Oliveira}, \citenamefont {Tcherbakoff},
  \citenamefont {Ruchon}, \citenamefont {Min\'ar},\ and\ \citenamefont
  {Hricovini}}]{Fanciulli2020}%
  \BibitemOpen
  \bibfield  {author} {\bibinfo {author} {\bibfnamefont {M.}~\bibnamefont
  {Fanciulli}}, \bibinfo {author} {\bibfnamefont {J.}~\bibnamefont {Schusser}},
  \bibinfo {author} {\bibfnamefont {M.-I.}\ \bibnamefont {Lee}}, \bibinfo
  {author} {\bibfnamefont {Z.~E.}\ \bibnamefont {Youbi}}, \bibinfo {author}
  {\bibfnamefont {O.}~\bibnamefont {Heckmann}}, \bibinfo {author}
  {\bibfnamefont {M.~C.}\ \bibnamefont {Richter}}, \bibinfo {author}
  {\bibfnamefont {C.}~\bibnamefont {Cacho}}, \bibinfo {author} {\bibfnamefont
  {C.}~\bibnamefont {Spezzani}}, \bibinfo {author} {\bibfnamefont
  {D.}~\bibnamefont {Bresteau}}, \bibinfo {author} {\bibfnamefont {J.-F.
  m.~c.}\ \bibnamefont {Hergott}}, \bibinfo {author} {\bibfnamefont
  {P.}~\bibnamefont {D'Oliveira}}, \bibinfo {author} {\bibfnamefont
  {O.}~\bibnamefont {Tcherbakoff}}, \bibinfo {author} {\bibfnamefont
  {T.}~\bibnamefont {Ruchon}}, \bibinfo {author} {\bibfnamefont
  {J.}~\bibnamefont {Min\'ar}},\ and\ \bibinfo {author} {\bibfnamefont
  {K.}~\bibnamefont {Hricovini}},\ }\bibfield  {title} {\bibinfo {title} {Spin,
  time, and angle resolved photoemission spectroscopy on
  ${\mathrm{wte}}_{2}$},\ }\href
  {https://doi.org/10.1103/PhysRevResearch.2.013261} {\bibfield  {journal}
  {\bibinfo  {journal} {Phys. Rev. Research}\ }\textbf {\bibinfo {volume}
  {2}},\ \bibinfo {pages} {013261} (\bibinfo {year} {2020})}\BibitemShut
  {NoStop}%
\bibitem [{\citenamefont {Wan}\ \emph {et~al.}(2022)\citenamefont {Wan},
  \citenamefont {Wang}, \citenamefont {Kuroda}, \citenamefont {Zhang},
  \citenamefont {Koshiishi}, \citenamefont {Suzuki}, \citenamefont {Kim},
  \citenamefont {Noguchi}, \citenamefont {Bareille}, \citenamefont {Yaji},
  \citenamefont {Harasawa}, \citenamefont {Shin}, \citenamefont {Cheong},
  \citenamefont {Fujimori},\ and\ \citenamefont {Kondo}}]{Wan2022}%
  \BibitemOpen
  \bibfield  {author} {\bibinfo {author} {\bibfnamefont {Y.}~\bibnamefont
  {Wan}}, \bibinfo {author} {\bibfnamefont {L.}~\bibnamefont {Wang}}, \bibinfo
  {author} {\bibfnamefont {K.}~\bibnamefont {Kuroda}}, \bibinfo {author}
  {\bibfnamefont {P.}~\bibnamefont {Zhang}}, \bibinfo {author} {\bibfnamefont
  {K.}~\bibnamefont {Koshiishi}}, \bibinfo {author} {\bibfnamefont
  {M.}~\bibnamefont {Suzuki}}, \bibinfo {author} {\bibfnamefont
  {J.}~\bibnamefont {Kim}}, \bibinfo {author} {\bibfnamefont {R.}~\bibnamefont
  {Noguchi}}, \bibinfo {author} {\bibfnamefont {C.}~\bibnamefont {Bareille}},
  \bibinfo {author} {\bibfnamefont {K.}~\bibnamefont {Yaji}}, \bibinfo {author}
  {\bibfnamefont {A.}~\bibnamefont {Harasawa}}, \bibinfo {author}
  {\bibfnamefont {S.}~\bibnamefont {Shin}}, \bibinfo {author} {\bibfnamefont
  {S.-W.}\ \bibnamefont {Cheong}}, \bibinfo {author} {\bibfnamefont
  {A.}~\bibnamefont {Fujimori}},\ and\ \bibinfo {author} {\bibfnamefont
  {T.}~\bibnamefont {Kondo}},\ }\bibfield  {title} {\bibinfo {title} {Selective
  observation of surface and bulk bands in polar ${\mathrm{wte}}_{2}$ by
  laser-based spin- and angle-resolved photoemission spectroscopy},\ }\href
  {https://doi.org/10.1103/PhysRevB.105.085421} {\bibfield  {journal} {\bibinfo
   {journal} {Phys. Rev. B}\ }\textbf {\bibinfo {volume} {105}},\ \bibinfo
  {pages} {085421} (\bibinfo {year} {2022})}\BibitemShut {NoStop}%
\bibitem [{\citenamefont {Escher}\ \emph {et~al.}(2011)\citenamefont {Escher},
  \citenamefont {Weber}, \citenamefont {Merkel}, \citenamefont {Plucinski},\
  and\ \citenamefont {Schneider}}]{Escher2011}%
  \BibitemOpen
  \bibfield  {author} {\bibinfo {author} {\bibfnamefont {M.}~\bibnamefont
  {Escher}}, \bibinfo {author} {\bibfnamefont {N.~B.}\ \bibnamefont {Weber}},
  \bibinfo {author} {\bibfnamefont {M.}~\bibnamefont {Merkel}}, \bibinfo
  {author} {\bibfnamefont {L.}~\bibnamefont {Plucinski}},\ and\ \bibinfo
  {author} {\bibfnamefont {C.~M.}\ \bibnamefont {Schneider}},\ }\bibfield
  {title} {\bibinfo {title} {Ferrum: A new highly efficient spin detector for
  electron spectroscopy},\ }\href@noop {} {\bibfield  {journal} {\bibinfo
  {journal} {e-J. Surf. Sci. Nanotech.}\ }\textbf {\bibinfo {volume} {9}},\
  \bibinfo {pages} {340} (\bibinfo {year} {2011})}\BibitemShut {NoStop}%
\bibitem [{\citenamefont {Okuda}\ \emph {et~al.}(2008)\citenamefont {Okuda},
  \citenamefont {Takeichi}, \citenamefont {Maeda}, \citenamefont {Harasawa},
  \citenamefont {Matsuda}, \citenamefont {Kinoshita},\ and\ \citenamefont
  {Kakizaki}}]{Okuda2008}%
  \BibitemOpen
  \bibfield  {author} {\bibinfo {author} {\bibfnamefont {T.}~\bibnamefont
  {Okuda}}, \bibinfo {author} {\bibfnamefont {Y.}~\bibnamefont {Takeichi}},
  \bibinfo {author} {\bibfnamefont {Y.}~\bibnamefont {Maeda}}, \bibinfo
  {author} {\bibfnamefont {A.}~\bibnamefont {Harasawa}}, \bibinfo {author}
  {\bibfnamefont {I.}~\bibnamefont {Matsuda}}, \bibinfo {author} {\bibfnamefont
  {T.}~\bibnamefont {Kinoshita}},\ and\ \bibinfo {author} {\bibfnamefont
  {A.}~\bibnamefont {Kakizaki}},\ }\bibfield  {title} {\bibinfo {title} {A new
  spin-polarized photoemission spectrometer with very high efficiency and
  energy resolution},\ }\href {https://doi.org/10.1063/1.3058757} {\bibfield
  {journal} {\bibinfo  {journal} {Review of Scientific Instruments}\ }\textbf
  {\bibinfo {volume} {79}},\ \bibinfo {pages} {123117} (\bibinfo {year}
  {2008})},\ \Eprint {https://arxiv.org/abs/https://doi.org/10.1063/1.3058757}
  {https://doi.org/10.1063/1.3058757} \BibitemShut {NoStop}%
\bibitem [{\citenamefont {Bigi}\ \emph {et~al.}(2017)\citenamefont {Bigi},
  \citenamefont {Das}, \citenamefont {Benedetti}, \citenamefont {Salvador},
  \citenamefont {Krizmancic}, \citenamefont {Sergo}, \citenamefont {Martin},
  \citenamefont {Panaccione}, \citenamefont {Rossi}, \citenamefont {Fujii},\
  and\ \citenamefont {Vobornik}}]{Bigi2017}%
  \BibitemOpen
  \bibfield  {author} {\bibinfo {author} {\bibfnamefont {C.}~\bibnamefont
  {Bigi}}, \bibinfo {author} {\bibfnamefont {P.~K.}\ \bibnamefont {Das}},
  \bibinfo {author} {\bibfnamefont {D.}~\bibnamefont {Benedetti}}, \bibinfo
  {author} {\bibfnamefont {F.}~\bibnamefont {Salvador}}, \bibinfo {author}
  {\bibfnamefont {D.}~\bibnamefont {Krizmancic}}, \bibinfo {author}
  {\bibfnamefont {R.}~\bibnamefont {Sergo}}, \bibinfo {author} {\bibfnamefont
  {A.}~\bibnamefont {Martin}}, \bibinfo {author} {\bibfnamefont
  {G.}~\bibnamefont {Panaccione}}, \bibinfo {author} {\bibfnamefont
  {G.}~\bibnamefont {Rossi}}, \bibinfo {author} {\bibfnamefont
  {J.}~\bibnamefont {Fujii}},\ and\ \bibinfo {author} {\bibfnamefont
  {I.}~\bibnamefont {Vobornik}},\ }\bibfield  {title} {\bibinfo {title} {{Very
  efficient spin polarization analysis (VESPA): new exchange scattering-based
  setup for spin-resolved ARPES at APE-NFFA beamline at Elettra}},\ }\href
  {https://doi.org/10.1107/S1600577517006907} {\bibfield  {journal} {\bibinfo
  {journal} {Journal of Synchrotron Radiation}\ }\textbf {\bibinfo {volume}
  {24}},\ \bibinfo {pages} {750} (\bibinfo {year} {2017})}\BibitemShut
  {NoStop}%
\bibitem [{\citenamefont {Perdew}\ \emph {et~al.}(1996)\citenamefont {Perdew},
  \citenamefont {Burke},\ and\ \citenamefont {Ernzerhof}}]{Perdew1996}%
  \BibitemOpen
  \bibfield  {author} {\bibinfo {author} {\bibfnamefont {J.~P.}\ \bibnamefont
  {Perdew}}, \bibinfo {author} {\bibfnamefont {K.}~\bibnamefont {Burke}},\ and\
  \bibinfo {author} {\bibfnamefont {M.}~\bibnamefont {Ernzerhof}},\ }\bibfield
  {title} {\bibinfo {title} {Generalized gradient approximation made simple},\
  }\href {https://doi.org/10.1103/PhysRevLett.77.3865} {\bibfield  {journal}
  {\bibinfo  {journal} {Phys. Rev. Lett.}\ }\textbf {\bibinfo {volume} {77}},\
  \bibinfo {pages} {3865} (\bibinfo {year} {1996})}\BibitemShut {NoStop}%
\bibitem [{\citenamefont {FLEUR}()}]{Fleur}%
  \BibitemOpen
  \bibfield  {author} {\bibinfo {author} {\bibnamefont {FLEUR}},\ }\href@noop
  {} {}\bibinfo {howpublished} {\url{http://www.flapw.de/}}\BibitemShut
  {NoStop}%
\bibitem [{\citenamefont {Braun}\ \emph {et~al.}(2018)\citenamefont {Braun},
  \citenamefont {Minár},\ and\ \citenamefont {Ebert}}]{Braun2018}%
  \BibitemOpen
  \bibfield  {author} {\bibinfo {author} {\bibfnamefont {J.}~\bibnamefont
  {Braun}}, \bibinfo {author} {\bibfnamefont {J.}~\bibnamefont {Minár}},\ and\
  \bibinfo {author} {\bibfnamefont {H.}~\bibnamefont {Ebert}},\ }\bibfield
  {title} {\bibinfo {title} {Correlation, temperature and disorder: Recent
  developments in the one-step description of angle-resolved photoemission},\
  }\href {https://doi.org/https://doi.org/10.1016/j.physrep.2018.02.007}
  {\bibfield  {journal} {\bibinfo  {journal} {Physics Reports}\ }\textbf
  {\bibinfo {volume} {740}},\ \bibinfo {pages} {1} (\bibinfo {year} {2018})},\
  \bibinfo {note} {correlation, temperature and disorder: Recent developments
  in the one-step description of angle-resolved photoemission}\BibitemShut
  {NoStop}%
\bibitem [{\citenamefont {Weber}\ \emph {et~al.}(2018)\citenamefont {Weber},
  \citenamefont {R\"u\ss{}mann}, \citenamefont {Xu}, \citenamefont {Muff},
  \citenamefont {Fanciulli}, \citenamefont {Magrez}, \citenamefont {Bugnon},
  \citenamefont {Berger}, \citenamefont {Plumb}, \citenamefont {Shi},
  \citenamefont {Bl\"ugel}, \citenamefont {Mavropoulos},\ and\ \citenamefont
  {Dil}}]{Weber2018}%
  \BibitemOpen
  \bibfield  {author} {\bibinfo {author} {\bibfnamefont {A.~P.}\ \bibnamefont
  {Weber}}, \bibinfo {author} {\bibfnamefont {P.}~\bibnamefont
  {R\"u\ss{}mann}}, \bibinfo {author} {\bibfnamefont {N.}~\bibnamefont {Xu}},
  \bibinfo {author} {\bibfnamefont {S.}~\bibnamefont {Muff}}, \bibinfo {author}
  {\bibfnamefont {M.}~\bibnamefont {Fanciulli}}, \bibinfo {author}
  {\bibfnamefont {A.}~\bibnamefont {Magrez}}, \bibinfo {author} {\bibfnamefont
  {P.}~\bibnamefont {Bugnon}}, \bibinfo {author} {\bibfnamefont
  {H.}~\bibnamefont {Berger}}, \bibinfo {author} {\bibfnamefont {N.~C.}\
  \bibnamefont {Plumb}}, \bibinfo {author} {\bibfnamefont {M.}~\bibnamefont
  {Shi}}, \bibinfo {author} {\bibfnamefont {S.}~\bibnamefont {Bl\"ugel}},
  \bibinfo {author} {\bibfnamefont {P.}~\bibnamefont {Mavropoulos}},\ and\
  \bibinfo {author} {\bibfnamefont {J.~H.}\ \bibnamefont {Dil}},\ }\bibfield
  {title} {\bibinfo {title} {Spin-resolved electronic response to the phase
  transition in ${\mathrm{mote}}_{2}$},\ }\href
  {https://doi.org/10.1103/PhysRevLett.121.156401} {\bibfield  {journal}
  {\bibinfo  {journal} {Phys. Rev. Lett.}\ }\textbf {\bibinfo {volume} {121}},\
  \bibinfo {pages} {156401} (\bibinfo {year} {2018})}\BibitemShut {NoStop}%
\bibitem [{\citenamefont {Moser}(2017)}]{Moser2017}%
  \BibitemOpen
  \bibfield  {author} {\bibinfo {author} {\bibfnamefont {S.}~\bibnamefont
  {Moser}},\ }\bibfield  {title} {\bibinfo {title} {An experimentalist's guide
  to the matrix element in angle resolved photoemission},\ }\href
  {https://doi.org/https://doi.org/10.1016/j.elspec.2016.11.007} {\bibfield
  {journal} {\bibinfo  {journal} {Journal of Electron Spectroscopy and Related
  Phenomena}\ }\textbf {\bibinfo {volume} {214}},\ \bibinfo {pages} {29 }
  (\bibinfo {year} {2017})}\BibitemShut {NoStop}%
\bibitem [{\citenamefont {Zhu}\ \emph {et~al.}(2013)\citenamefont {Zhu},
  \citenamefont {Veenstra}, \citenamefont {Levy}, \citenamefont {Ubaldini},
  \citenamefont {Syers}, \citenamefont {Butch}, \citenamefont {Paglione},
  \citenamefont {Haverkort}, \citenamefont {Elfimov},\ and\ \citenamefont
  {Damascelli}}]{Zhu2013}%
  \BibitemOpen
  \bibfield  {author} {\bibinfo {author} {\bibfnamefont {Z.-H.}\ \bibnamefont
  {Zhu}}, \bibinfo {author} {\bibfnamefont {C.~N.}\ \bibnamefont {Veenstra}},
  \bibinfo {author} {\bibfnamefont {G.}~\bibnamefont {Levy}}, \bibinfo {author}
  {\bibfnamefont {A.}~\bibnamefont {Ubaldini}}, \bibinfo {author}
  {\bibfnamefont {P.}~\bibnamefont {Syers}}, \bibinfo {author} {\bibfnamefont
  {N.~P.}\ \bibnamefont {Butch}}, \bibinfo {author} {\bibfnamefont
  {J.}~\bibnamefont {Paglione}}, \bibinfo {author} {\bibfnamefont {M.~W.}\
  \bibnamefont {Haverkort}}, \bibinfo {author} {\bibfnamefont {I.~S.}\
  \bibnamefont {Elfimov}},\ and\ \bibinfo {author} {\bibfnamefont
  {A.}~\bibnamefont {Damascelli}},\ }\bibfield  {title} {\bibinfo {title}
  {Layer-by-layer entangled spin-orbital texture of the topological surface
  state in ${\mathrm{bi}}_{2}{\mathrm{se}}_{3}$},\ }\href
  {https://doi.org/10.1103/PhysRevLett.110.216401} {\bibfield  {journal}
  {\bibinfo  {journal} {Phys. Rev. Lett.}\ }\textbf {\bibinfo {volume} {110}},\
  \bibinfo {pages} {216401} (\bibinfo {year} {2013})}\BibitemShut {NoStop}%
\bibitem [{\citenamefont {Xiao}\ \emph {et~al.}(2020)\citenamefont {Xiao},
  \citenamefont {Wang}, \citenamefont {Wang}, \citenamefont {Pemmaraju},
  \citenamefont {Wang}, \citenamefont {Muscher}, \citenamefont {Sie},
  \citenamefont {Nyby}, \citenamefont {Devereaux}, \citenamefont {Qian},
  \citenamefont {Zhang},\ and\ \citenamefont {Lindenberg}}]{Xiao2020}%
  \BibitemOpen
  \bibfield  {author} {\bibinfo {author} {\bibfnamefont {J.}~\bibnamefont
  {Xiao}}, \bibinfo {author} {\bibfnamefont {Y.}~\bibnamefont {Wang}}, \bibinfo
  {author} {\bibfnamefont {H.}~\bibnamefont {Wang}}, \bibinfo {author}
  {\bibfnamefont {C.~D.}\ \bibnamefont {Pemmaraju}}, \bibinfo {author}
  {\bibfnamefont {S.}~\bibnamefont {Wang}}, \bibinfo {author} {\bibfnamefont
  {P.}~\bibnamefont {Muscher}}, \bibinfo {author} {\bibfnamefont {E.~J.}\
  \bibnamefont {Sie}}, \bibinfo {author} {\bibfnamefont {C.~M.}\ \bibnamefont
  {Nyby}}, \bibinfo {author} {\bibfnamefont {T.~P.}\ \bibnamefont {Devereaux}},
  \bibinfo {author} {\bibfnamefont {X.}~\bibnamefont {Qian}}, \bibinfo {author}
  {\bibfnamefont {X.}~\bibnamefont {Zhang}},\ and\ \bibinfo {author}
  {\bibfnamefont {A.~M.}\ \bibnamefont {Lindenberg}},\ }\bibfield  {title}
  {\bibinfo {title} {Berry curvature memory through electrically driven
  stacking transitions},\ }\href {https://doi.org/10.1038/s41567-020-0947-0}
  {\bibfield  {journal} {\bibinfo  {journal} {Nature Physics}\ }\textbf
  {\bibinfo {volume} {16}},\ \bibinfo {pages} {1028} (\bibinfo {year}
  {2020})}\BibitemShut {NoStop}%
\bibitem [{\citenamefont {Tusche}\ \emph {et~al.}(2015)\citenamefont {Tusche},
  \citenamefont {Krasyuk},\ and\ \citenamefont {Kirschner}}]{Tusche2015}%
  \BibitemOpen
  \bibfield  {author} {\bibinfo {author} {\bibfnamefont {C.}~\bibnamefont
  {Tusche}}, \bibinfo {author} {\bibfnamefont {A.}~\bibnamefont {Krasyuk}},\
  and\ \bibinfo {author} {\bibfnamefont {J.}~\bibnamefont {Kirschner}},\
  }\bibfield  {title} {\bibinfo {title} {Spin resolved bandstructure imaging
  with a high resolution momentum microscope},\ }\href
  {https://doi.org/https://doi.org/10.1016/j.ultramic.2015.03.020} {\bibfield
  {journal} {\bibinfo  {journal} {Ultramicroscopy}\ }\textbf {\bibinfo {volume}
  {159}},\ \bibinfo {pages} {520} (\bibinfo {year} {2015})},\ \bibinfo {note}
  {special Issue: LEEM-PEEM 9}\BibitemShut {NoStop}%
\bibitem [{\citenamefont {Schönhense}\ \emph {et~al.}(2015)\citenamefont
  {Schönhense}, \citenamefont {Medjanik},\ and\ \citenamefont
  {Elmers}}]{Schoenhense2015}%
  \BibitemOpen
  \bibfield  {author} {\bibinfo {author} {\bibfnamefont {G.}~\bibnamefont
  {Schönhense}}, \bibinfo {author} {\bibfnamefont {K.}~\bibnamefont
  {Medjanik}},\ and\ \bibinfo {author} {\bibfnamefont {H.-J.}\ \bibnamefont
  {Elmers}},\ }\bibfield  {title} {\bibinfo {title} {Space-, time- and
  spin-resolved photoemission},\ }\href
  {https://doi.org/https://doi.org/10.1016/j.elspec.2015.05.016} {\bibfield
  {journal} {\bibinfo  {journal} {Journal of Electron Spectroscopy and Related
  Phenomena}\ }\textbf {\bibinfo {volume} {200}},\ \bibinfo {pages} {94}
  (\bibinfo {year} {2015})},\ \bibinfo {note} {special Anniversary Issue:
  Volume 200}\BibitemShut {NoStop}%
\bibitem [{\citenamefont {Bentmann}\ \emph {et~al.}(2017)\citenamefont
  {Bentmann}, \citenamefont {Maa\ss{}}, \citenamefont {Krasovskii},
  \citenamefont {Peixoto}, \citenamefont {Seibel}, \citenamefont {Leandersson},
  \citenamefont {Balasubramanian},\ and\ \citenamefont
  {Reinert}}]{Bentmann2017}%
  \BibitemOpen
  \bibfield  {author} {\bibinfo {author} {\bibfnamefont {H.}~\bibnamefont
  {Bentmann}}, \bibinfo {author} {\bibfnamefont {H.}~\bibnamefont {Maa\ss{}}},
  \bibinfo {author} {\bibfnamefont {E.~E.}\ \bibnamefont {Krasovskii}},
  \bibinfo {author} {\bibfnamefont {T.~R.~F.}\ \bibnamefont {Peixoto}},
  \bibinfo {author} {\bibfnamefont {C.}~\bibnamefont {Seibel}}, \bibinfo
  {author} {\bibfnamefont {M.}~\bibnamefont {Leandersson}}, \bibinfo {author}
  {\bibfnamefont {T.}~\bibnamefont {Balasubramanian}},\ and\ \bibinfo {author}
  {\bibfnamefont {F.}~\bibnamefont {Reinert}},\ }\bibfield  {title} {\bibinfo
  {title} {Strong linear dichroism in spin-polarized photoemission from
  spin-orbit-coupled surface states},\ }\href
  {https://doi.org/10.1103/PhysRevLett.119.106401} {\bibfield  {journal}
  {\bibinfo  {journal} {Phys. Rev. Lett.}\ }\textbf {\bibinfo {volume} {119}},\
  \bibinfo {pages} {106401} (\bibinfo {year} {2017})}\BibitemShut {NoStop}%
\bibitem [{\citenamefont {Yaji}\ \emph {et~al.}(2017)\citenamefont {Yaji},
  \citenamefont {Kuroda}, \citenamefont {Toyohisa}, \citenamefont {Harasawa},
  \citenamefont {Ishida}, \citenamefont {Watanabe}, \citenamefont {Chen},
  \citenamefont {Kobayashi}, \citenamefont {Komori},\ and\ \citenamefont
  {Shin}}]{Yaji2017}%
  \BibitemOpen
  \bibfield  {author} {\bibinfo {author} {\bibfnamefont {K.}~\bibnamefont
  {Yaji}}, \bibinfo {author} {\bibfnamefont {K.}~\bibnamefont {Kuroda}},
  \bibinfo {author} {\bibfnamefont {S.}~\bibnamefont {Toyohisa}}, \bibinfo
  {author} {\bibfnamefont {A.}~\bibnamefont {Harasawa}}, \bibinfo {author}
  {\bibfnamefont {Y.}~\bibnamefont {Ishida}}, \bibinfo {author} {\bibfnamefont
  {S.}~\bibnamefont {Watanabe}}, \bibinfo {author} {\bibfnamefont
  {C.}~\bibnamefont {Chen}}, \bibinfo {author} {\bibfnamefont {K.}~\bibnamefont
  {Kobayashi}}, \bibinfo {author} {\bibfnamefont {F.}~\bibnamefont {Komori}},\
  and\ \bibinfo {author} {\bibfnamefont {S.}~\bibnamefont {Shin}},\ }\bibfield
  {title} {\bibinfo {title} {Spin-dependent quantum interference in
  photoemission process from spin-orbit coupled states},\ }\bibfield  {journal}
  {\bibinfo  {journal} {Nature Communications}\ }\textbf {\bibinfo {volume}
  {8}},\ \href {https://doi.org/10.1038/ncomms14588} {10.1038/ncomms14588}
  (\bibinfo {year} {2017})\BibitemShut {NoStop}%
\bibitem [{\citenamefont {Henk}\ \emph {et~al.}(2003)\citenamefont {Henk},
  \citenamefont {Bose}, \citenamefont {Michael},\ and\ \citenamefont
  {Bruno}}]{Henk2003}%
  \BibitemOpen
  \bibfield  {author} {\bibinfo {author} {\bibfnamefont {J.}~\bibnamefont
  {Henk}}, \bibinfo {author} {\bibfnamefont {P.}~\bibnamefont {Bose}}, \bibinfo
  {author} {\bibfnamefont {T.}~\bibnamefont {Michael}},\ and\ \bibinfo {author}
  {\bibfnamefont {P.}~\bibnamefont {Bruno}},\ }\bibfield  {title} {\bibinfo
  {title} {Spin motion of photoelectrons},\ }\href
  {https://doi.org/10.1103/PhysRevB.68.052403} {\bibfield  {journal} {\bibinfo
  {journal} {Phys. Rev. B}\ }\textbf {\bibinfo {volume} {68}},\ \bibinfo
  {pages} {052403} (\bibinfo {year} {2003})}\BibitemShut {NoStop}%
\bibitem [{\citenamefont {Berdot}\ \emph {et~al.}(2010)\citenamefont {Berdot},
  \citenamefont {Hallal}, \citenamefont {Bismaths}, \citenamefont {Joly},
  \citenamefont {Dey}, \citenamefont {Henk}, \citenamefont {Alouani},\ and\
  \citenamefont {Weber}}]{Berdot2010}%
  \BibitemOpen
  \bibfield  {author} {\bibinfo {author} {\bibfnamefont {T.}~\bibnamefont
  {Berdot}}, \bibinfo {author} {\bibfnamefont {A.}~\bibnamefont {Hallal}},
  \bibinfo {author} {\bibfnamefont {L.~T.}\ \bibnamefont {Bismaths}}, \bibinfo
  {author} {\bibfnamefont {L.}~\bibnamefont {Joly}}, \bibinfo {author}
  {\bibfnamefont {P.}~\bibnamefont {Dey}}, \bibinfo {author} {\bibfnamefont
  {J.}~\bibnamefont {Henk}}, \bibinfo {author} {\bibfnamefont {M.}~\bibnamefont
  {Alouani}},\ and\ \bibinfo {author} {\bibfnamefont {W.}~\bibnamefont
  {Weber}},\ }\bibfield  {title} {\bibinfo {title} {Effect of submonolayer mgo
  coverages on the electron-spin motion in fe(001): Experiment and theory},\
  }\href {https://doi.org/10.1103/PhysRevB.82.172407} {\bibfield  {journal}
  {\bibinfo  {journal} {Phys. Rev. B}\ }\textbf {\bibinfo {volume} {82}},\
  \bibinfo {pages} {172407} (\bibinfo {year} {2010})}\BibitemShut {NoStop}%
\end{thebibliography}

%

\end{document}